\documentclass[twocolumn,trackchanges]{aastex701}
\usepackage{amsmath}
\usepackage{graphicx}
\usepackage{indentfirst}
\usepackage{float}%定义图片
\usepackage{graphicx}
\usepackage{subcaption}
\usepackage{caption}
\usepackage{multirow}
\usepackage{longtable}
\usepackage{booktabs}

\begin{document}

%\title{Brightest GRB flare: "double tracking" phenomenon in GRB flare}

%\title{Double-tracking behavior in the brightest GRB flare}

\title{Comprehensive Measurement of Spectral Evolution in a GRB Flare: High Time-Resolution Insights into the ``Double-Tracking" Phenomenon}

\correspondingauthor{Wen-Jun Tan, Shao-Lin Xiong}
\email{tanwj@ihep.ac.cn, xiongsl@ihep.ac.cn}

\author[0009-0002-6411-8422]{Zheng-Hang Yu}
\affil{State Key Laboratory of Particle Astrophysics, Institute of High Energy Physics, Chinese Academy of Sciences, 19B Yuquan Road, Beijing 100049, China}
\affil{University of Chinese Academy of Sciences, Chinese Academy of Sciences, Beijing 100049, China}
\email{zhyu@ihep.ac.cn}  
%\affiliation{University of Chinese Academy of Sciences, Chinese Academy of Sciences, Beijing 100049, China}

\author[0009-0006-5506-5970]{Wen-Jun Tan*}
\affil{State Key Laboratory of Particle Astrophysics, Institute of High Energy Physics, Chinese Academy of Sciences, 19B Yuquan Road, Beijing 100049, China}
\affil{University of Chinese Academy of Sciences, Chinese Academy of Sciences, Beijing 100049, China}
\email{}

\author[0009-0008-8053-2985]{Chen-Wei Wang}
\affil{State Key Laboratory of Particle Astrophysics, Institute of High Energy Physics, Chinese Academy of Sciences, 19B Yuquan Road, Beijing 100049, China}
\affil{University of Chinese Academy of Sciences, Chinese Academy of Sciences, Beijing 100049, China}
\email{}

\author[0000-0002-4771-7653]{Shao-Lin Xiong*} 
\affil{State Key Laboratory of Particle Astrophysics, Institute of High Energy Physics, Chinese Academy of Sciences, 19B Yuquan Road, Beijing 100049, China}
\email{}

\author[0009-0001-7226-2355]{Chao Zheng}
\affil{State Key Laboratory of Particle Astrophysics, Institute of High Energy Physics, Chinese Academy of Sciences, 19B Yuquan Road, Beijing 100049, China}
\affil{TIANFU Cosmic Ray Research Center, Chengdu, Sichuan 610000, China}
\email{}

%%%%%%%%%%%%%%%%%%%%%%%%%%%%%%%%%%

  \author[0000-0002-8097-3616]{Peng Zhang}
 \affil{State Key Laboratory of Particle Astrophysics, Institute of High Energy Physics, Chinese Academy of Sciences, 19B Yuquan Road, Beijing 100049, China}
 \affil{College of Electronic and Information Engineering, Tongji University, Shanghai 201804, China}
 \email{}

\author[]{Zheng-Hua An}
\affil{State Key Laboratory of Particle Astrophysics, Institute of High Energy Physics, Chinese Academy of Sciences, 19B Yuquan Road, Beijing 100049, China}
 \email{}

\author{Ce Cai}
\affil{College of Physics and Hebei Key Laboratory of Photophysics Research and Application, 
Hebei Normal University, Shijiazhuang, Hebei 050024, China}
\email{}

\author[]{Min Gao}
\affil{State Key Laboratory of Particle Astrophysics, Institute of High Energy Physics, Chinese Academy of Sciences, 19B Yuquan Road, Beijing 100049, China}
 \email{}

\author[]{Ke Gong}
\affil{State Key Laboratory of Particle Astrophysics, Institute of High Energy Physics, Chinese Academy of Sciences, 19B Yuquan Road, Beijing 100049, China}
 \email{}

 \author[]{Dong-Ya Guo}
\affil{State Key Laboratory of Particle Astrophysics, Institute of High Energy Physics, Chinese Academy of Sciences, 19B Yuquan Road, Beijing 100049, China}
 \email{}

 \author[]{Hao-Xuan Guo}
 \affil{State Key Laboratory of Particle Astrophysics, Institute of High Energy Physics, Chinese Academy of Sciences, 19B Yuquan Road, Beijing 100049, China}
 \affil{Department of Nuclear Science and Technology, School of Energy and Power Engineering, Xi'an Jiaotong University, Xi'an, China}
 \email{}

\author[]{Yue Huang}
\affil{State Key Laboratory of Particle Astrophysics, Institute of High Energy Physics, Chinese Academy of Sciences, 19B Yuquan Road, Beijing 100049, China}
 \email{}

\author[]{Bing Li}
\affil{State Key Laboratory of Particle Astrophysics, Institute of High Energy Physics, Chinese Academy of Sciences, 19B Yuquan Road, Beijing 100049, China}
 \email{}
 
\author[]{Cheng-Kui Li}
\affil{State Key Laboratory of Particle Astrophysics, Institute of High Energy Physics, Chinese Academy of Sciences, 19B Yuquan Road, Beijing 100049, China}
 \email{}

\author[]{Xiao-Bo Li}
\affil{State Key Laboratory of Particle Astrophysics, Institute of High Energy Physics, Chinese Academy of Sciences, 19B Yuquan Road, Beijing 100049, China}
\email{}

\author[]{Xin-Qiao Li}
\affil{State Key Laboratory of Particle Astrophysics, Institute of High Energy Physics, Chinese Academy of Sciences, 19B Yuquan Road, Beijing 100049, China}
 \email{}

  \author[0009-0004-1887-4686]{Jia-Cong Liu}
 \affil{State Key Laboratory of Particle Astrophysics, Institute of High Energy Physics, Chinese Academy of Sciences, 19B Yuquan Road, Beijing 100049, China}
 \affil{University of Chinese Academy of Sciences, Chinese Academy of Sciences, Beijing 100049, China}
 \email{}

 \author[]{Ya-Qing Liu}
\affil{State Key Laboratory of Particle Astrophysics, Institute of High Energy Physics, Chinese Academy of Sciences, 19B Yuquan Road, Beijing 100049, China}
 \email{}

\author[]{Xiao-Jing Liu}
\affil{State Key Laboratory of Particle Astrophysics, Institute of High Energy Physics, Chinese Academy of Sciences, 19B Yuquan Road, Beijing 100049, China}
 \email{}

\author[]{Xiang Ma}
\affil{State Key Laboratory of Particle Astrophysics, Institute of High Energy Physics, Chinese Academy of Sciences, 19B Yuquan Road, Beijing 100049, China}
 \email{}

 \author[]{Wen-Xi Peng}
\affil{State Key Laboratory of Particle Astrophysics, Institute of High Energy Physics, Chinese Academy of Sciences, 19B Yuquan Road, Beijing 100049, China}
 \email{}

 \author[]{Rui Qiao}
\affil{State Key Laboratory of Particle Astrophysics, Institute of High Energy Physics, Chinese Academy of Sciences, 19B Yuquan Road, Beijing 100049, China}
 \email{}

  \author[]{Yang-Zhao Ren}
 \affil{State Key Laboratory of Particle Astrophysics, Institute of High Energy Physics, Chinese Academy of Sciences, 19B Yuquan Road, Beijing 100049, China}
 \affil{School of Physical Science and Technology, Southwest Jiaotong University, Chengdu 611756, China}
 \email{}

\author[]{Li-Ming Song}
\affil{State Key Laboratory of Particle Astrophysics, Institute of High Energy Physics, Chinese Academy of Sciences, 19B Yuquan Road, Beijing 100049, China}
\affil{University of Chinese Academy of Sciences, Chinese Academy of Sciences, Beijing 100049, China}
\email{}

\author[]{Jin Wang}
\affil{State Key Laboratory of Particle Astrophysics, Institute of High Energy Physics, Chinese Academy of Sciences, 19B Yuquan Road, Beijing 100049, China}
 \email{}

 \author[]{Jin-Zhou Wang}
\affil{State Key Laboratory of Particle Astrophysics, Institute of High Energy Physics, Chinese Academy of Sciences, 19B Yuquan Road, Beijing 100049, China}
 \email{}

\author[]{Ping Wang}
\affil{State Key Laboratory of Particle Astrophysics, Institute of High Energy Physics, Chinese Academy of Sciences, 19B Yuquan Road, Beijing 100049, China}
 \email{}

  \author[0009-0008-5068-3504]{Yue Wang}
 \affil{State Key Laboratory of Particle Astrophysics, Institute of High Energy Physics, Chinese Academy of Sciences, 19B Yuquan Road, Beijing 100049, China}
 \affil{University of Chinese Academy of Sciences, Chinese Academy of Sciences, Beijing 100049, China}
 \email{}

\author[]{Xiang-Yang Wen}
\affil{State Key Laboratory of Particle Astrophysics, Institute of High Energy Physics, Chinese Academy of Sciences, 19B Yuquan Road, Beijing 100049, China}
\email{}

\author{Shuo Xiao}
\affil{School of Physics and Electronic Science, Guizhou Normal University, Guiyang 550001, China}
\affil{Guizhou Provincial Key Laboratory of Radio Astronomy and Data Processing, Guizhou Normal University, Guiyang 550001, China}
\email{}

 \author[0000-0001-9217-7070]{Sheng-Lun Xie} 
 \affil{State Key Laboratory of Particle Astrophysics, Institute of High Energy Physics, Chinese Academy of Sciences, 19B Yuquan Road, Beijing 100049, China}
 \affil{Institute of Astrophysics, Central China Normal University, Wuhan 430079, China}
 \email{}

 \author[0000-0001-8664-5085]{Wang-Chen Xue}
 \affil{State Key Laboratory of Particle Astrophysics, Institute of High Energy Physics, Chinese Academy of Sciences, 19B Yuquan Road, Beijing 100049, China}
 \affil{University of Chinese Academy of Sciences, Chinese Academy of Sciences, Beijing 100049, China}
 \email{}

\author[]{Sheng Yang}
\affil{State Key Laboratory of Particle Astrophysics, Institute of High Energy Physics, Chinese Academy of Sciences, 19B Yuquan Road, Beijing 100049, China}
 \email{}
 
\author[]{Shu-Xu Yi}
\affil{State Key Laboratory of Particle Astrophysics, Institute of High Energy Physics, Chinese Academy of Sciences, 19B Yuquan Road, Beijing 100049, China}
 \email{}

%\author[]{Qi-Bin Yi}
%\affil{State Key Laboratory of Particle Astrophysics, Institute of High Energy Physics, Chinese Academy of Sciences, 19B Yuquan Road, Beijing 100049, China}
%\affil{School of Physics and Optoelectronics, Xiangtan University, Xiangtan 411105, China}
% \email{}

\author[]{Da-Li Zhang}
\affil{State Key Laboratory of Particle Astrophysics, Institute of High Energy Physics, Chinese Academy of Sciences, 19B Yuquan Road, Beijing 100049, China}
 \email{}

\author[]{Fan Zhang}
\affil{State Key Laboratory of Particle Astrophysics, Institute of High Energy Physics, Chinese Academy of Sciences, 19B Yuquan Road, Beijing 100049, China}
 \email{}

\author[]{Zhen Zhang}
\affil{State Key Laboratory of Particle Astrophysics, Institute of High Energy Physics, Chinese Academy of Sciences, 19B Yuquan Road, Beijing 100049, China}
 \email{}

\author[]{Xiao-Yun Zhao}
\affil{State Key Laboratory of Particle Astrophysics, Institute of High Energy Physics, Chinese Academy of Sciences, 19B Yuquan Road, Beijing 100049, China}
 \email{}

%\author[]{Yi Zhao}
%\affil{School of Computer and Information, Dezhou University, Dezhou 253023, Shandong, China}
% \email{}

\author[]{Jin-Peng Zhang}
 \affil{State Key Laboratory of Particle Astrophysics, Institute of High Energy Physics, Chinese Academy of Sciences, 19B Yuquan Road, Beijing 100049, China}
 \affil{University of Chinese Academy of Sciences, Chinese Academy of Sciences, Beijing 100049, China}
 \email{}

 \author[0009-0008-6247-0645]{Wen-Long Zhang}
\affil{Purple Mountain Observatory, Chinese Academy of Sciences, Nanjing 210023, China}
\affil{School of Astronomy and Space Sciences, University of Science and Technology of China, Hefei 230026, China}
\email{wlzhang@pmo.ac.cn}

\author[0000-0001-5348-7033]{Yan-Qiu Zhang}
\affil{School of Physics and Electronic Science, Guizhou Normal University, Guiyang 550001, China}
\affil{Guizhou Provincial Key Laboratory of Radio Astronomy and Data Processing, Guizhou Normal University, Guiyang 550001, China}
\email{}

\author[]{Shi-Jie Zheng}
\affil{State Key Laboratory of Particle Astrophysics, Institute of High Energy Physics, Chinese Academy of Sciences, 19B Yuquan Road, Beijing 100049, China}
 \email{}

\author[]{Shuang-Nan Zhang}
\affil{State Key Laboratory of Particle Astrophysics, Institute of High Energy Physics, Chinese Academy of Sciences, 19B Yuquan Road, Beijing 100049, China}
\affil{University of Chinese Academy of Sciences, Chinese Academy of Sciences, Beijing 100049, China}
\email{}

\begin{abstract}

The spectral evolution characteristics of the prompt emission in gamma-ray bursts (GRBs) have been extensively studied, but detailed investigations of spectral evolution in a GRB flare remain lacking. In this work, we present the first analysis of spectral parameter evolution in a GRB flare through high time-resolved spectral fitting of the Brightest Flare in GRB 221009A. We find that the $\alpha$-$Flux$, $E_\text{p}$-$Flux$, and $E_\text{p}$-$\alpha$ relationships during both the overall phase and the rise phase of flare can be well described by simple power-law model, showing positive correlations. Therefore, we conclude that Brightest Flare exhibits ``Double-tracking" behavior. Since values of $\alpha$ do not exceed the synchrotron ``death line" (-2/3), we explain this phenomenon using a magnetic dissipation synchrotron radiation model. In the decay phase of flare, the $E_\text{p}$–$Flux$ and $E_\text{p}$–$\alpha$ correlations become notably flatter, with their power-law indices decreasing significantly compared to those in the rise phase. This may be due to the fact that the next flare begins to erupt before the Brightest Flare has completely ended, resulting in the combined effects of both two flares.  Our study of spectral parameter relations of the Brightest Flare provides new insights into the radiation mechanisms of both GRB prompt emission and flares.

\end{abstract}

\keywords{Gamma-Ray Bursts}

\section{Introduction} 

Gamma-ray bursts (GRBs) are the most energetic and violent astrophysical phenomena in the universe \citep{2018_grb_physics_zhang_bing,2025_Eiso,2025_GRB,2026_0702}. GRB radiation could be divided into two phases: prompt emission and afterglow emission. Flares are often observed during the afterglow phase. Observations from the \textit{Swift} satellite indicate that about half of the GRBs have X-ray flares \citep{A1_2006_flare_physics, 2018_grb_physics_zhang_bing}. These flares can occur at any phase during the afterglow \citep{A2_2011_flare_every}, and a single GRB may exhibit one or multiple flares. Previous studies favor a common physical origin between X-ray flares and prompt emission and also indicate that flares most likely originate from late activities of the central engine \citep{A1_2006_flare_physics,A3_2007_flare_statis1,A4_2007_flare_statis2,C111_Yi_2016_x_ray_flare,09A_flare}.

The time-resolved or time-integrated spectra of most observed GRBs are non-thermal and can be fitted using the Band function \citep{A5_Band} or the cutoff power-law (CPL) model \citep{A6_CPL}. The low-energy spectral index $\alpha$ and the peak energy in the $\nu F_{\nu}$ spectrum $E_{\text{p}}$ are key parameters derived from GRB spectral fitting. The prevailing theory suggests that these non-thermal spectra may originate from synchrotron radiation or synchrotron self-Compton scattering \citep{A7_GRB_model_1,A8_GRB_model_2,A9_GRB_model_3}. The evolution of $\alpha$ and $E_{\text{p}}$ in the time-resolved spectra serves as an important probe to study the underlying physics of GRBs. Although the relations among $E_{\text{p}}$, $\alpha$, and flux ($F$) during the prompt emission have been extensively studied \citep{A12_para_relation_1,A10_EP_F,A11_alpha,A15_para_relation_4,A13_para_relation_2,A14_para_relation_3}, the spectral parameter relations in GRB flares remain unexplored.

In prompt emission of GRBs, $E_{\text{p}}$ is generally considered to exhibit two evolutionary patterns: (1) ``flux-tracking," also referred to as $E_{\text{p}}$-intensity, where it hardens when flux increases and softens when flux decreases \citep{A16_Ep_intensity_1,A17_Ep_intensity_2,A22_double_tracking}; (2) ``hard-to-soft," where the overall $E_{\text{p}}$ trend from the beginning to the end of the burst shows a progression from hard to soft \citep{A12_para_relation_1,A18_Ep_hard_soft}. Statistical analyzes from \textit{Fermi} and other satellite observations have confirmed these as the two dominant $E_{\text{p}}$ evolution patterns in GRBs, the former accounting for approximately one-third and the latter about two-thirds of cases \citep{A10_EP_F,A13_para_relation_2}.

Initially, the spectral index $\alpha$ in the prompt emission is assumed to remain constant. Under a steady magnetic field, the power-law index of the electron distribution is -2, and $\alpha$ remained at -1.5 \citep{A19_decayB_alpha_flux}. However, subsequent observations indicated that $\alpha$ varies continuously during a GRB. Further research revealed that $\alpha$ exhibits a ``flux-tracking" characteristic. \cite{A20_alpha_flux_heat} analyzed \textit{Fermi}/GBM-detected GRBs and found that $\alpha$ and flux conform to the relationship $F \propto e^{k \alpha}$, with $k \approx 3$, which can be explained by entropy dominated photosphere heating models. This ``flux-tracking" trend also exists in non-thermal radiation and can be well explained by a synchrotron radiation in a decaying magnetic field \citep{A19_decayB_alpha_flux}. Beyond this behavior, \cite{A21_broken_alpha_flux} discovered a broken $\alpha$-intensity relation in GRB 230307A, where a smoothly broken power-law provides a superior fit to the $\alpha$-$F$ relation. The reason for this phenomenon is the transition from a thermal dominated component to a non-thermal dominated component process in GRB 230307A \citep{A21_broken_alpha_flux}.

In the prompt emission, both $E_{\text{p}}$ and $\alpha$ exhibit a pattern of intensity tracking, also known as ``Double-tracking" \citep{A22_double_tracking}. However, this phenomenon has not been previously identified in GRB flares. GRB 221009A is the Brightest Of All Time (BOAT) GRB ever detected, which has a  flare component \citep{1_09A_Gecam}. The flare of GRB 221009A is composed of multiple superimposed flares, among which the brightest one is referred to as the ``Brightest Flare" \citep{09A_flare}. 
%Thanks to the extreme brightness of the Brightest Flare, we were able to performe high-time-resolution spectral analysis to study the relations between the spectral parameters of a GRB flare, revealing for the first time the characteristic "Double-tracking" pattern in GRB flares. 
In this paper, we perform the analysis of time-resolved spectral parameter evolution of the Brightest Flare in GRB 221009A, and report the ``Double-tracking" phenomenon in the Brightest Flare, which is the first observation of ``Double-tracking" in a GRB flare. Specifically, We find that the $\alpha$-$F$, $E_\text{p}$-$F$, and $E_\text{p}$-$\alpha$ during both the overall flare and the rise phase of flare show positive correlations, and can be well described by simple power-law model. Hereafter, we abbreviate ``Brightest Flare'' as BFL for ease of reference.

This paper is organized as follows. The observation of GRB 221009A, the fitting of time-resolved spectrum and the relation analysis between spectral parameters are presented in Section~\ref{section2}. The physical interpretation of these spectral relations are shown in Section~\ref{section3}.
%We discuss the implications of our findings in Section~\ref{section4}.
The summary is shown in Section~\ref{section4}.

\section{Observation and analysis} \label{section2}

\subsection{Observation} 

GECAM (Gravitational Wave High-energy Electromagnetic Counterpart All-sky Monitor) is a constellation of gamma-ray monitors designed to search for high energy electromagnetic counterparts (EMs) associated with gravitational waves (GWs) and fast radio bursts (FRBs). In addition to its primary scientific goal, GECAM is also capable of monitoring and locating gamma-ray bursts (GRBs), soft gamma-ray repeaters (SGRs), and other gamma-ray transient sources. 

\begin{figure*}
\centering
\begin{tabular}{cc}
    \begin{minipage}{0.45\textwidth}
        \includegraphics[width = \textwidth]{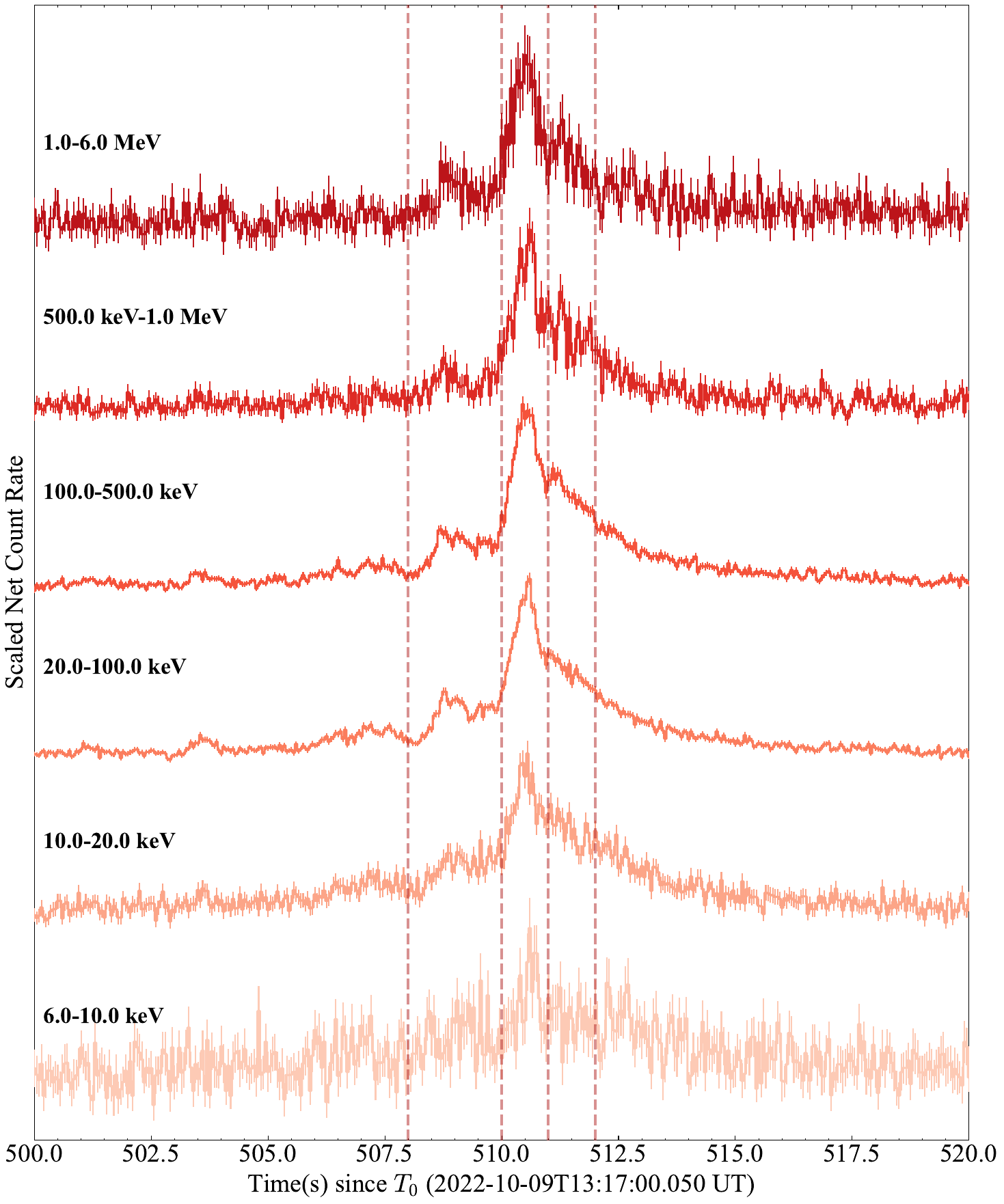}
        \subcaption{}
    \end{minipage}
    \begin{minipage}{0.42\textwidth}
        \includegraphics[width = \textwidth]{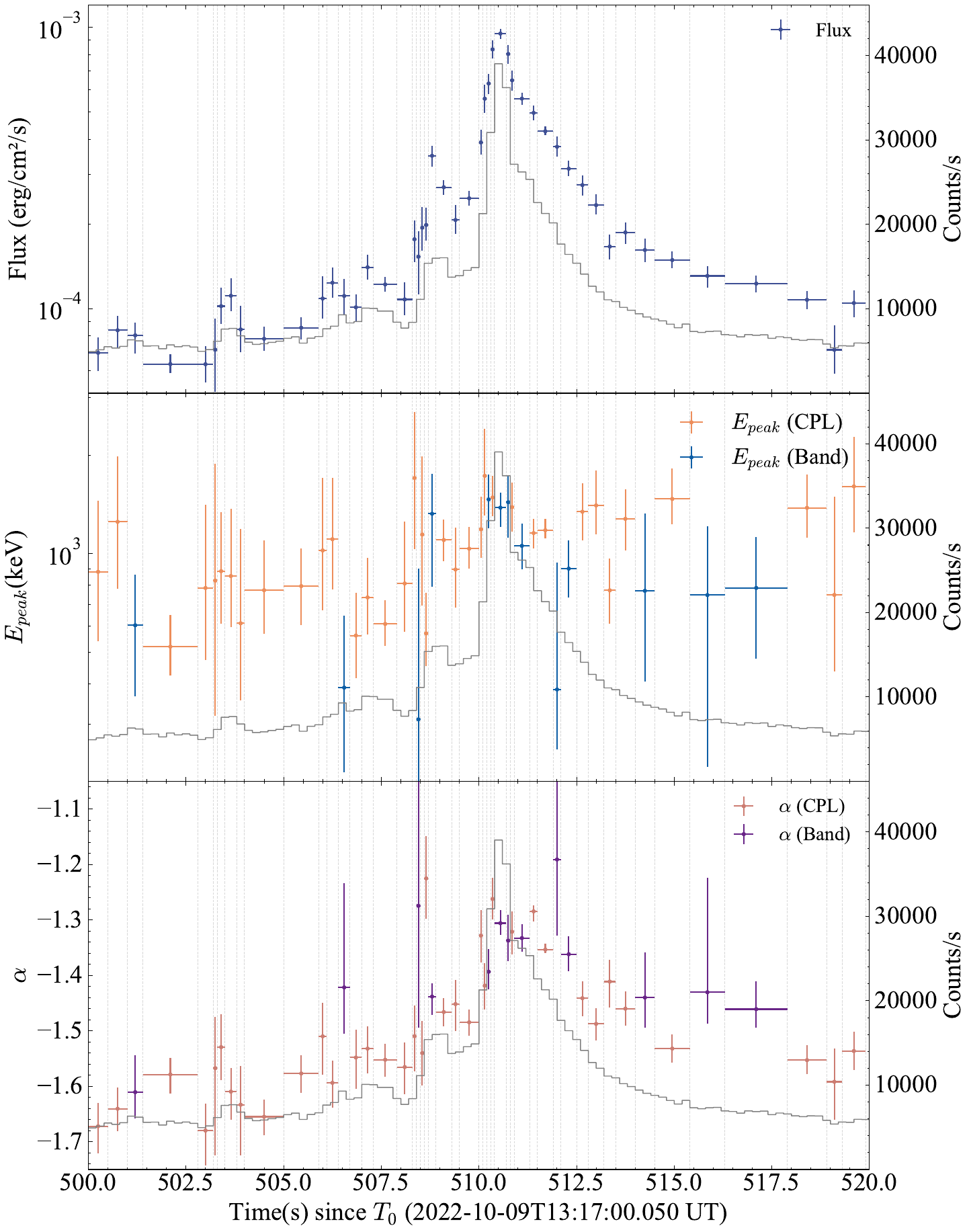}
        \subcaption{}
    \end{minipage}\\
    \begin{minipage}{0.8\textwidth}
        \includegraphics[width = \textwidth]{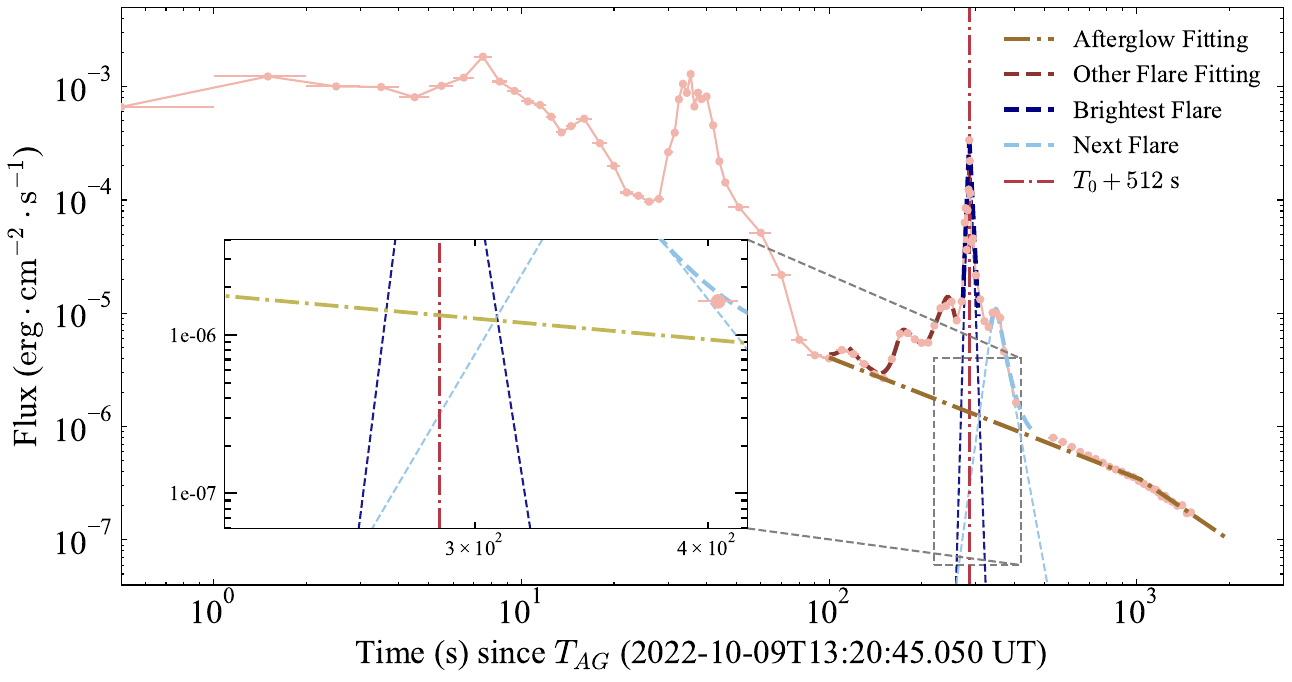}
        \subcaption{}
    \end{minipage}\\  
\end{tabular}
  \caption{\label{fig1}\small \textbf{Panel (a):} Light curves of BFL in different energy bands from keV to MeV. The vertical red dashed lines indicate the time-bin boundaries adopted from the S-2 scheme proposed by \cite{09A_flare}. A distinct high-energy pulse is observed between $T_0$+511\,s and $T_0$+512\,s. After $T_0$+512\,s, the light curve gradually transitions into a smooth decay. \textbf{Panel (b):} Evolution of the spectral parameters of the BFL. The three subplots, arranged from top to bottom, correspond to the evolution of flux, $E_{\text{p}}$, and $\alpha$. The $E_{\text{p}}$, and $\alpha$  calculated from different models are marked with distinct colors. \textbf{Panel (C):} Flux light curve of the GRB 221009A. The fitting parameters for the flare and afterglow are adopted from \cite{09A_flare}. The dark red vertical line marks the boundary within the decay phase, separating it into the pulse phase and the post-pulse phase. The dark blue and light blue dashed curves represent the fitting curves for the BFL and the flare following the BFL, respectively. Enlarged views of the boundary and the two fitted flares are shown on the left side of the figure.}
\end{figure*}

GECAM-C is equipped with a total of 12 gamma-ray detectors (GRDs) based on SiPM technology \citep{B9_2022_GRD} and 2 charged particle detectors (CPDs) made with readout plastic scintillators by SiPM array \citep{B10_2022_CPD}, with all detectors installed on two domes in the opposite sides of the satellite. 10 of 12 GRDs have two electron readout channels with different detection energy bands, referred to as high gain (HG) and low gain (LG), while the other 2 GRDs only have one readout channel. Previous studies have shown that GECAM-C has good temporal and spectral performance \citep{GECAM_callibration,B11_2024_ground_HEBs,B13_2024_calibration_GECAM}.

GRB 221009A is the Brightest Of All Time (BOAT) GRB ever detected. \textit{Fermi}/GBM first reported this event by real-time in-flight trigger \citep{B1_09A_Fermi,B8_09A_GBM}. \textit{Swift}/BAT provided the first accurate localization by observing the afterglow emission \citep{B2_09A_Swift}, eventually leading to the measurement of redshift of $z = 0.151$ \citep{B3_09A_redshift1, B4_09A_redshift2}. Remarkably, thousands of very high energy (VHE) photons (above 100 GeV) were detected by LHAASO from this burst \citep{B5_LHAASO_09A, B6_09A_soc}.

Specifically, \textit{Insight}-Hard X-ray Modulation Telescope (\textit{Insight}-HXMT) was triggered by GRB 221009A during its routine ground search at 13:17:00.050 UT on October 9, 2022, which is considered as the trigger time of GRB 221009A (denoted as $T_0$) in the current work \citep{1_09A_Gecam,B7_09A_insight}. Through a joint analysis of GRB 221009A with GECAM-C and \textit{Fermi}/GBM, a set of Gaussian emission lines following a power-law evolution has been identified, with the highest energy reaching 37\,MeV \citep{B15_2024_09A_line}.

GRB 221009A is exceptionally bright that most high energy telescopes (e.g. \textit{Fermi}/GBM, \textit{Insight}-HXMT) suffered instrumental effects (e.g., data saturation, pulse pileup) during its main burst epoch \citep[e.g.][]{B1_09A_Fermi, B7_09A_insight}. However, thanks to the dedicated designs for extremely bright bursts, GECAM-C made the uniquely accurate and high resolution measurements of the main burst, and found that this GRB has the largest $E_{\text{iso}} \sim 1.5 \times 10^{55}$ erg \citep{1_09A_Gecam}. 

GECAM-C successfully obtained complete observations of the GRB 221009A flare. During the detection of GRB 221009A, GECAM-C remained almost a constant pointing direction during both the burst and the revisit orbit periods, allowing us to use the data from the revisit orbit to estimate the background of GRB 221009A \citep{B14_2024_09A_afterglow}. In previous work, the revisit orbit method for background estimation has been proven to effectively measure the background of GRB 221009A \citep{1_09A_Gecam,B14_2024_09A_afterglow,B15_2024_09A_line}.

Previous studies of the flare associated with GRB 221009A have provided multiple observational lines of evidence indicating that the emission occurring after the main emission can be identified as GRB flare \citep{1_09A_Gecam,09A_flare}. 
%We discovered that this flare is composed of multiple superimposed flares \citep{09A_flare}. Among them, 
The BFL (from $T_0+500\,$\,s to $T_0+520$\,s) exhibits a record-breaking isotropic energy for GRB flares $E_{\text{iso}}= 1.82 \times 10^{53}\,erg$. 
%After performing a comprehensive integrated spectral fitting for BFL, we obtained the following parameters: $\alpha$ = -1.28, $\beta$ = -1.88, and $E_{\text{p}} = 300.91\,\rm keV$.
BFL possesses the highest $E_{\text{p}}$ observed in GRB flares so far. In the range of $1-10000\,\rm keV$, the fluence of the BFL reached $(3.49 \pm 0.05) \times 10^{-3}\,\rm{erg} \cdot \rm{cm}^{-2}$, exceeding that of most GRB prompt emission. Furthermore, we calculate a peak luminosity $L_{\text{peak}} = (4.51 \pm 0.12) \times 10^{52}\,\rm erg\cdot s^{-1}$ during the 1\,s peak time interval of the flare \citep{09A_flare}.

\subsection{Time-resolved spectral fit}

In this section, we conducted a more refined fitting and analysis of the time-resolved spectra. The Bayesian Block \citep{B16_BB} was used to segment the BFL into 49 time intervals. The specific time intervals is detailed in Table~\ref{table_spectrum}.

\begin{figure*}
  \centering
\begin{tabular}{ccc}
  \begin{minipage}{0.33\textwidth}
      \includegraphics[width = \textwidth]{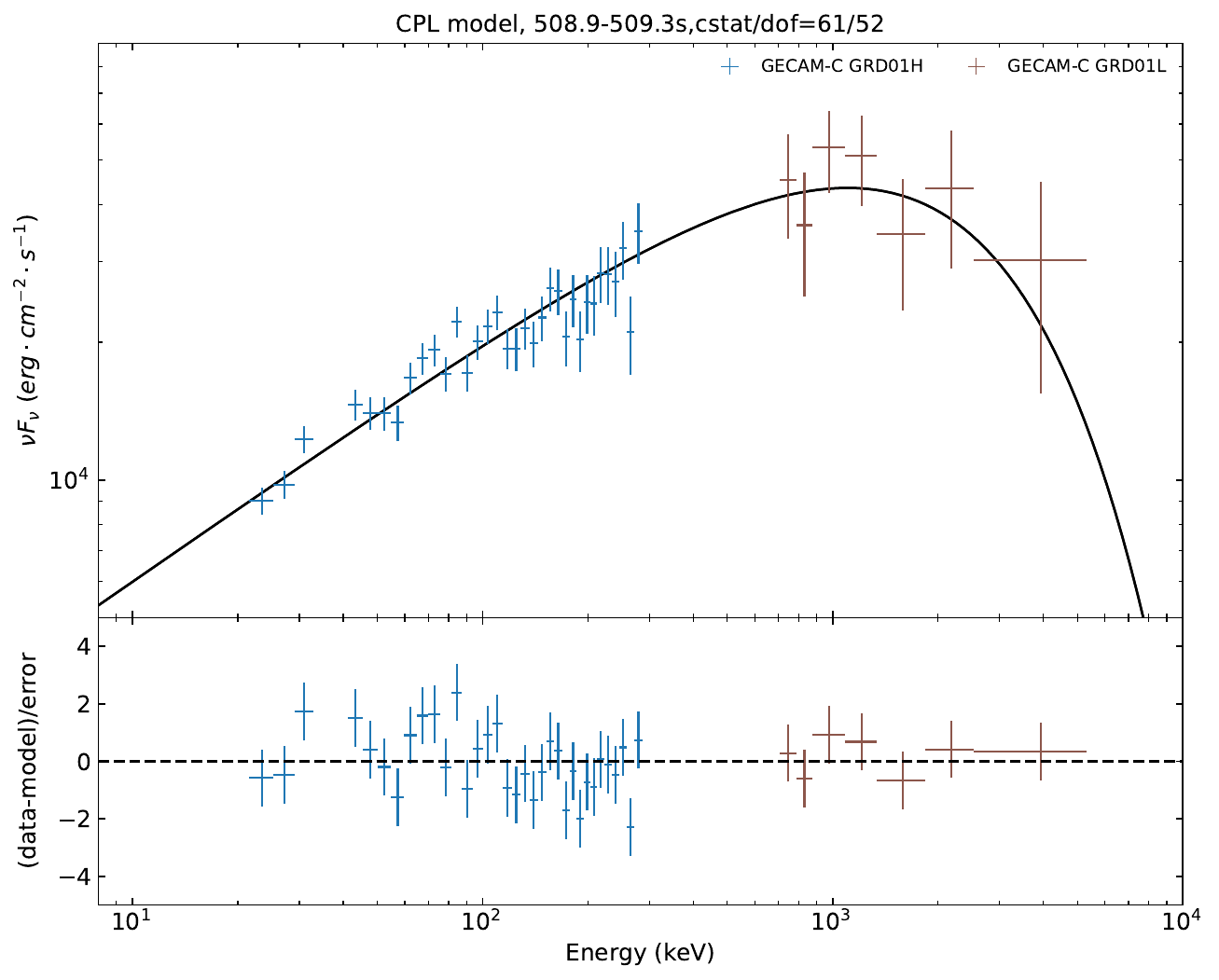}
      \subcaption{}
   \end{minipage}
    \begin{minipage}{0.33\textwidth}
        \includegraphics[width = \textwidth]{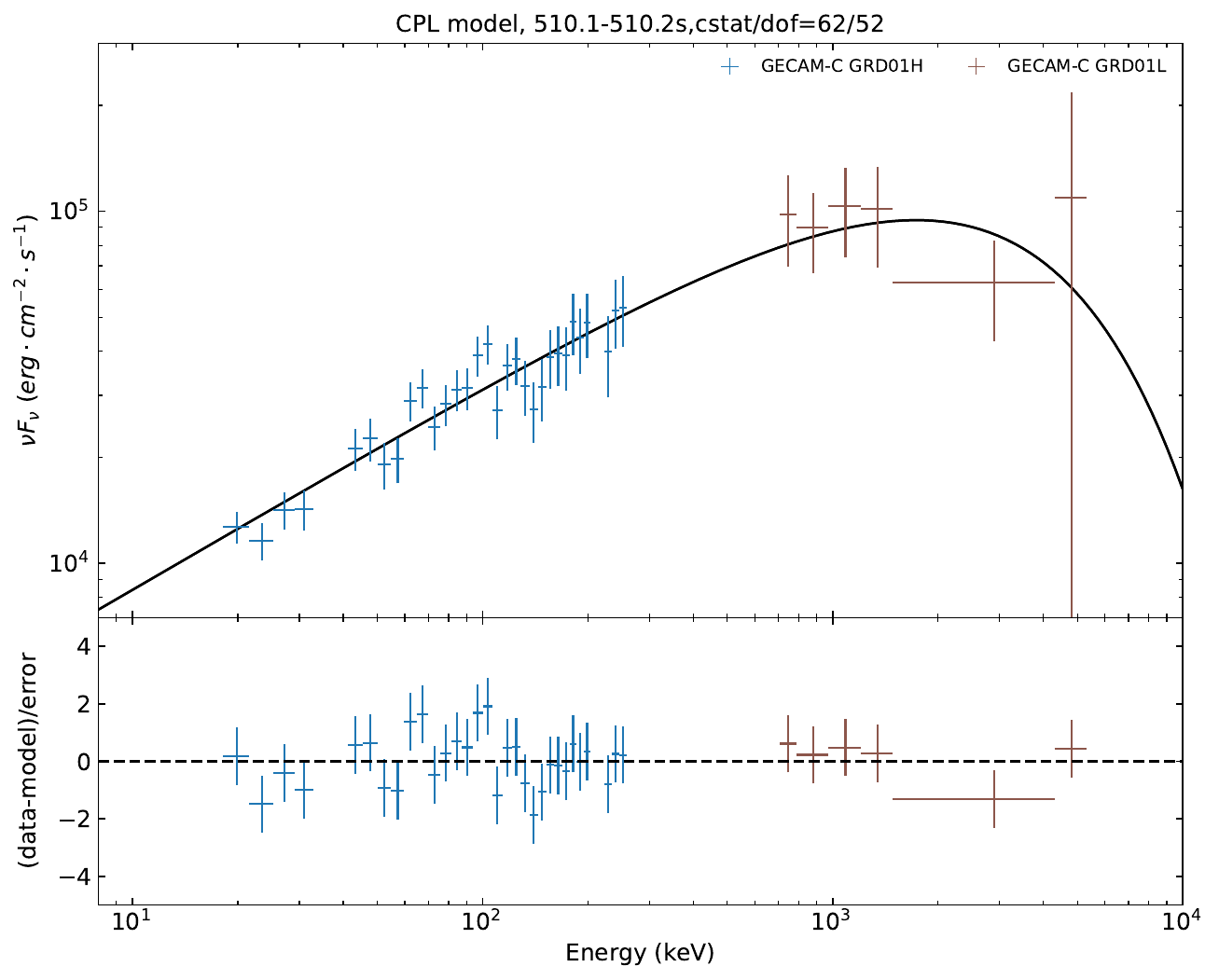}
        \subcaption{}
    \end{minipage}
    \begin{minipage}{0.33\textwidth}
        \includegraphics[width = \textwidth]{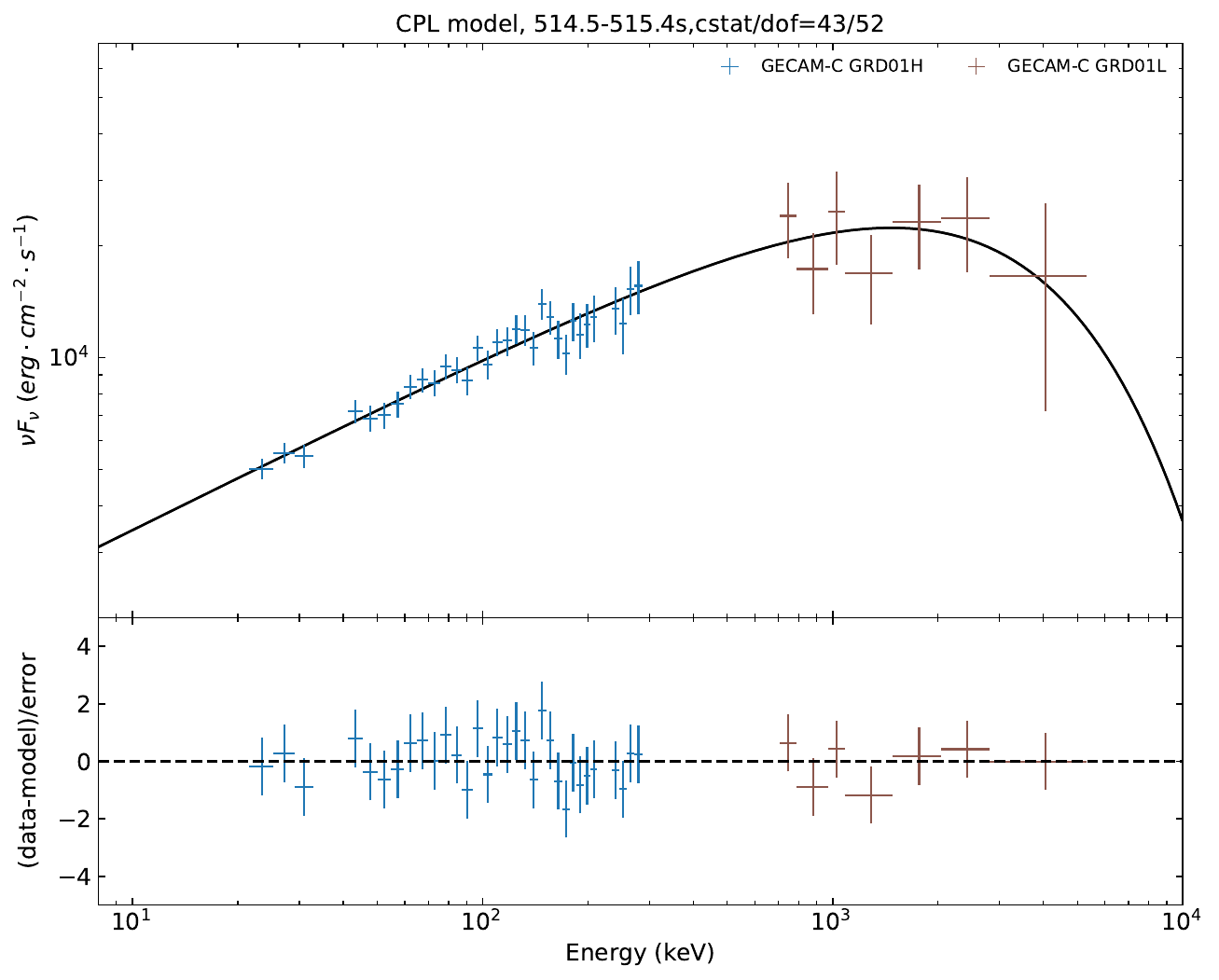}
        \subcaption{}
    \end{minipage}\\
    \begin{minipage}{0.33\textwidth}
        \includegraphics[width = \textwidth]{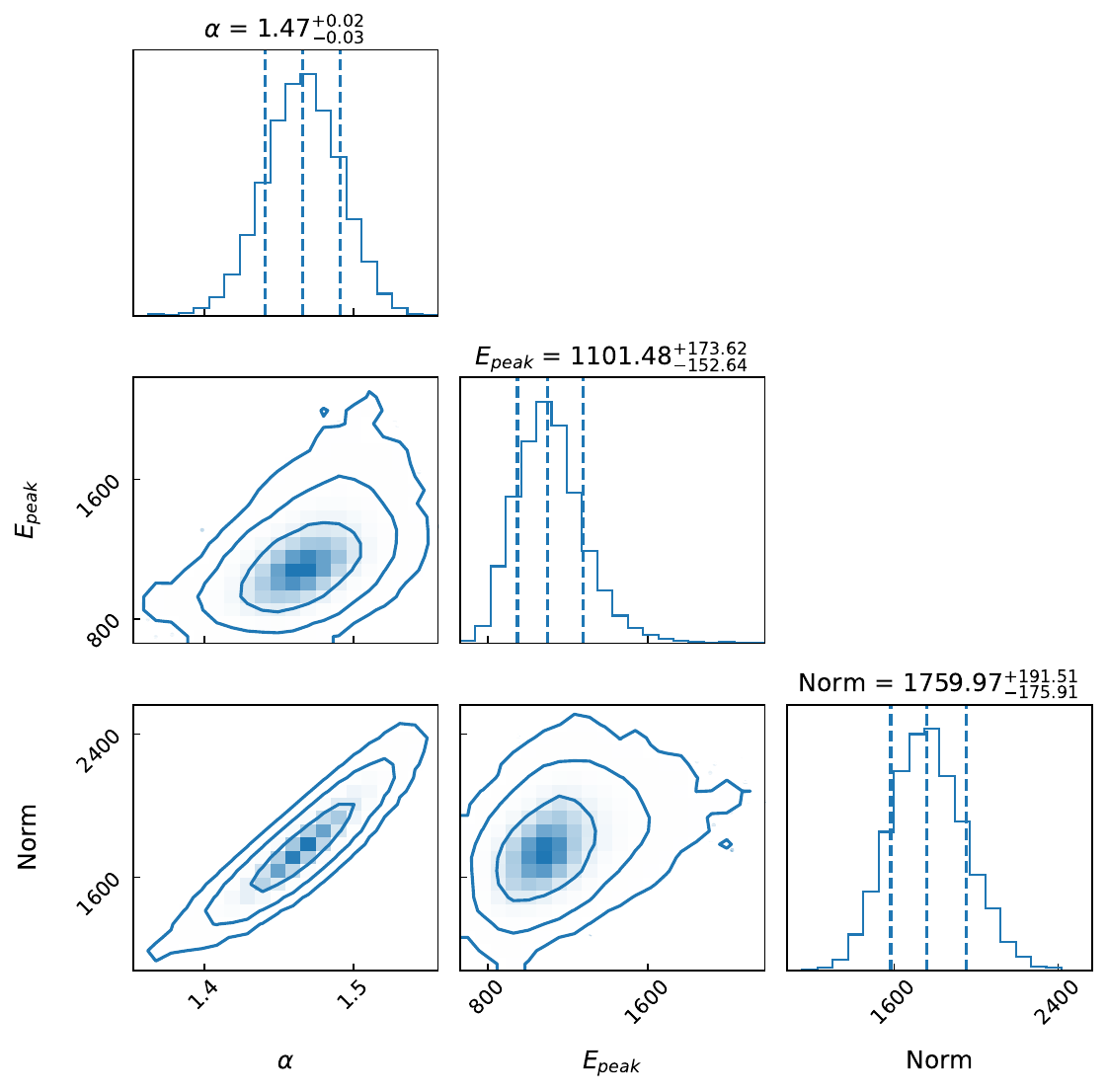}
        \subcaption{}
    \end{minipage}
    \begin{minipage}{0.33\textwidth}
        \includegraphics[width = \textwidth]{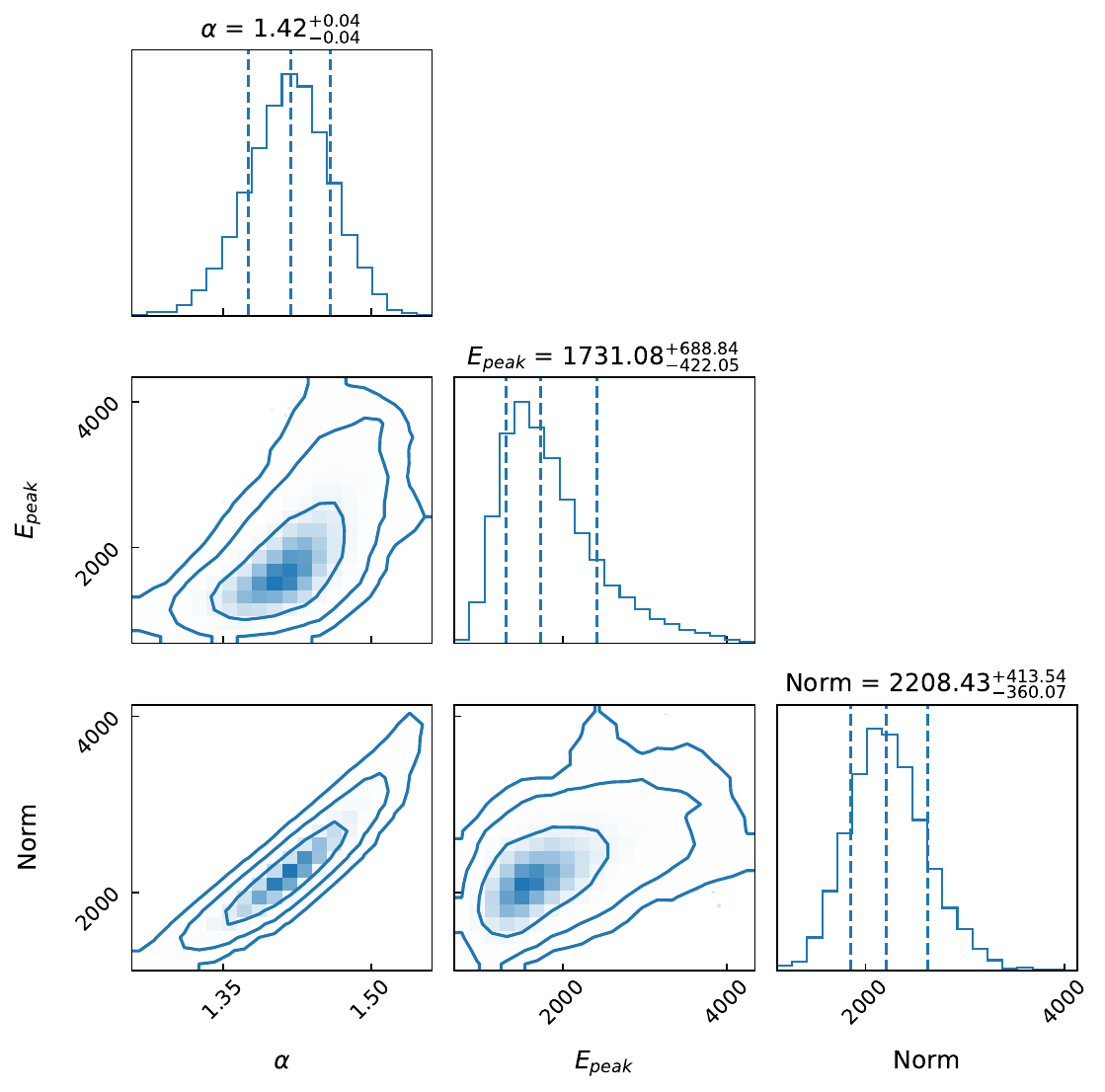}
        \subcaption{}
    \end{minipage}
    \begin{minipage}{0.33\textwidth}
        \includegraphics[width = \textwidth]{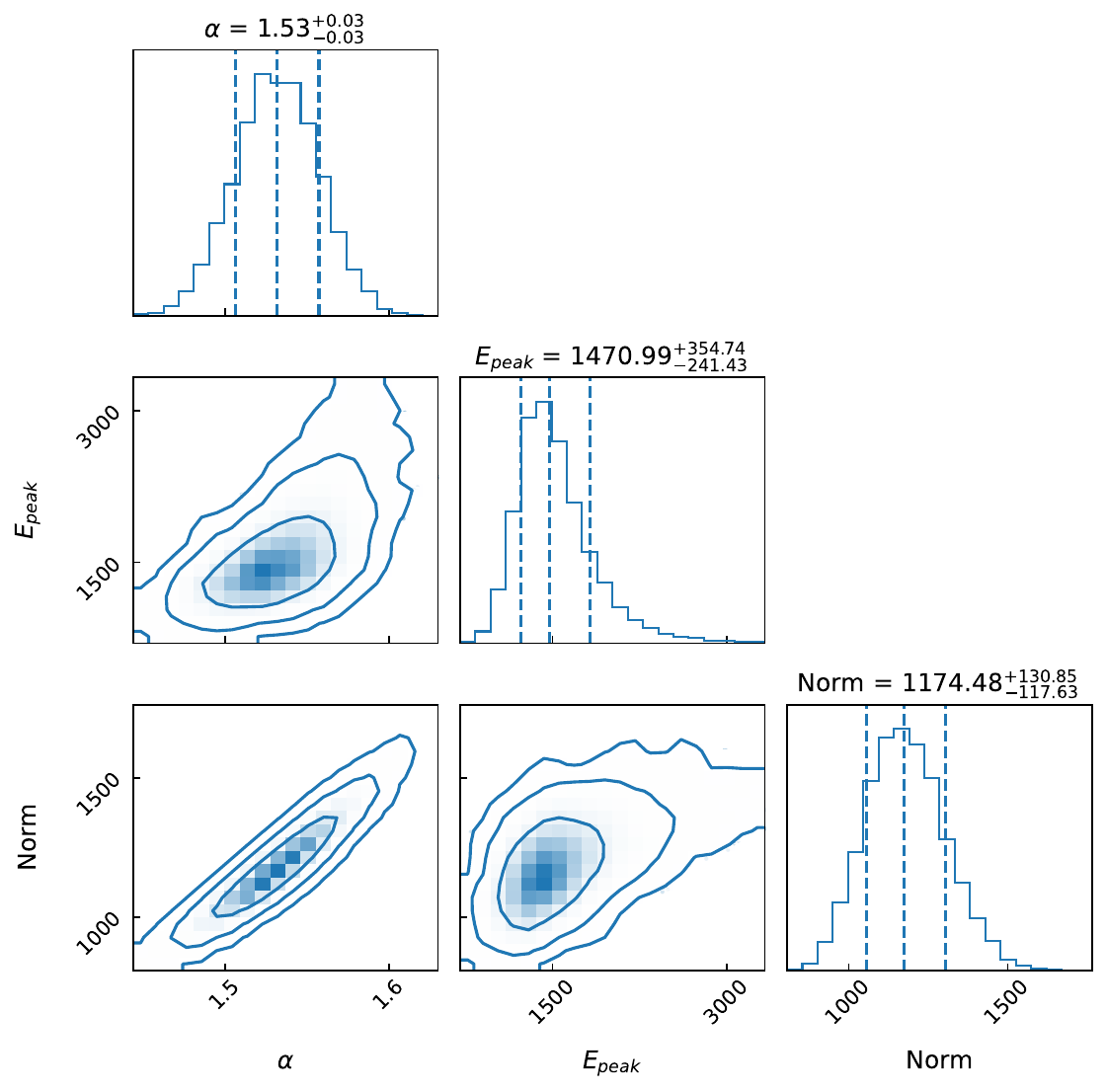}
        \subcaption{}
    \end{minipage}\\
\end{tabular}
  \caption{\label{fig_spectrum}\small Three energy spectrum fitting result with the CPL model and their corner plots. The energy spectrum fitting result graphs include the $\nu F_{\nu}$ plot and the residual maps. Corner plot showing the results of the spectral fitting. The contours represent the 1$\sigma$, 2$\sigma$, and 3$\sigma$ confidence levels.}
\end{figure*}

For spectral fitting of BFL, we primarily use the Band function \citep{A5_Band} and the CPL model \citep{A6_CPL}. The latest version of the GECAM-C CALDB and GECAMTools data analysis tools was used to generate the response matrix and spectrum files. Then, we used Xspec \citep{B17_1996_xspec} and PyXspec to fit the spectra of the flare. 

The Band and CPL are represented by Equations (\ref{eq1}) and (\ref{eq2}) respectively:

\begin{equation}
        N(E) =\begin{cases}
        A\left(\frac{E}{E_{\text{piv}}}\right)^{\alpha} \exp \left(-\frac{E}{E_{\text{c}}}\right), & E \leq(\alpha - \beta) E_{\text{c}}\\
        A\left[\frac{(\alpha - \beta) E_{\text{c}}}{E_{\text{piv}}}\right]^{\alpha - \beta} \exp (\beta - \alpha)\\\,\,\,\,\,\left(\frac{E}{E_{\text{piv}}}\right)^{\beta}, & E>(\alpha - \beta) E_{\text{c}} 
        \end{cases}\\\label{eq1}
\end{equation}

where $A$ is the normalization amplitude ($\rm photons \cdot cm^{-2} \cdot s^{-1} \cdot keV^{-1}$), $\alpha$ is the power law index \text{low-energy} and $\beta$ is the power law index \text{high-energy}, $E_{\rm c}$ is the characteristic energy in keV, $E_{\rm piv}$ is the pivot energy and usually fixed to 100 keV.

\begin{equation}
N(E) = A \left( \frac{E}{E_{\text{piv}}} \right)^\alpha \exp \left( -\frac{E}{E_{\text{c}}} \right) \label{eq2}
\end{equation}

where $A$ is the normalization amplitude ($\rm photons \cdot cm^{-2} \cdot s^{-1} \cdot keV^{-1}$); $\alpha$ is the power law photon index; $E_{\rm c}$ is the characteristic energy in keV; $E_{\rm piv}$ is 1 keV.

In our study of the BFL, we performed spectral analysis using the GECAM-C/GRD01 detector. The selected energy bands were 15--300\,keV for the HG and 0.7--5.5\,MeV for the LG \citep{GECAM_callibration}. We employed the C-statistic (\texttt{cstat}) method for the spectral fitting.

The time-resolved spectral fitting results for BFL are presented in Table~\ref{table_spectrum}. For both Band and CPL model, the peak energy $E_{\rm p}$ is related to $E_{\rm c}$ through $E_{\rm p}$ = $(2 + \alpha)E_{\rm c}$.
% Using the low-energy spectral index $\alpha$ and the break energy $E_{\text{cut}}$ obtained from the spectral fitting, the peak energy $E_{\text{p}}$ can be derived as $E_{\text{p}} = (\alpha + 2) E_{\text{cut}}$. 
The ratio of the c-statistic to the degrees of freedom (\texttt{cstat}/dof) is provided to assess the goodness of fit. The flux is calculated in 1-10000 keV. To compare and select the best model, we use the Bayesian Information Criterion (BIC) \citep{1978BIC}, which is defined as: $\text{BIC} = -2 \ln L + k \ln n$, where $L$ is the maximum likelihood value, $k$ represents the number of free parameters in the model, and $n$ is the number of data points.

\begin{figure*}[htbp]
  \centering
\begin{tabular}{ccC}
  \begin{minipage}{0.33\textwidth}
      \includegraphics[width = \textwidth]{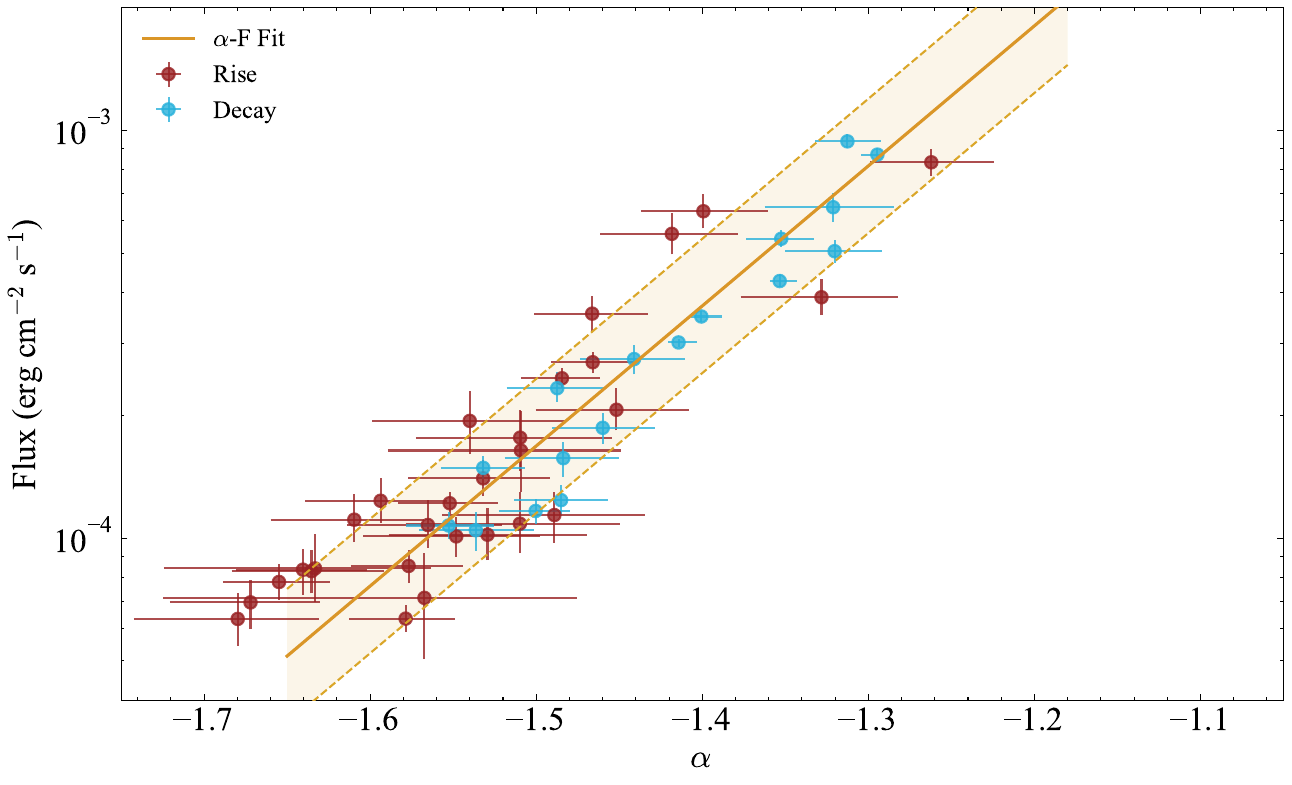}
      \subcaption{}
   \end{minipage}
    \begin{minipage}{0.33\textwidth}
        \includegraphics[width = \textwidth]{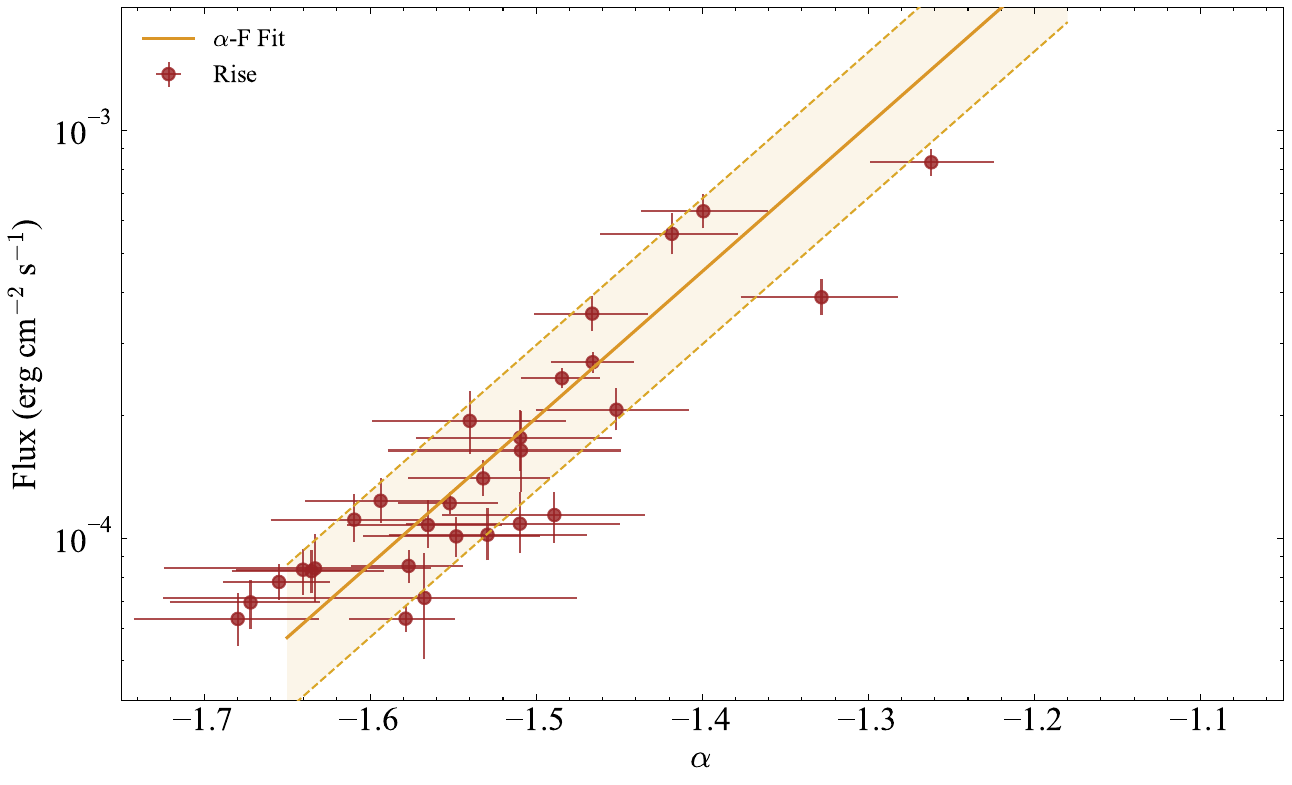}
        \subcaption{}
    \end{minipage}
    \begin{minipage}{0.33\textwidth}
        \includegraphics[width = \textwidth]{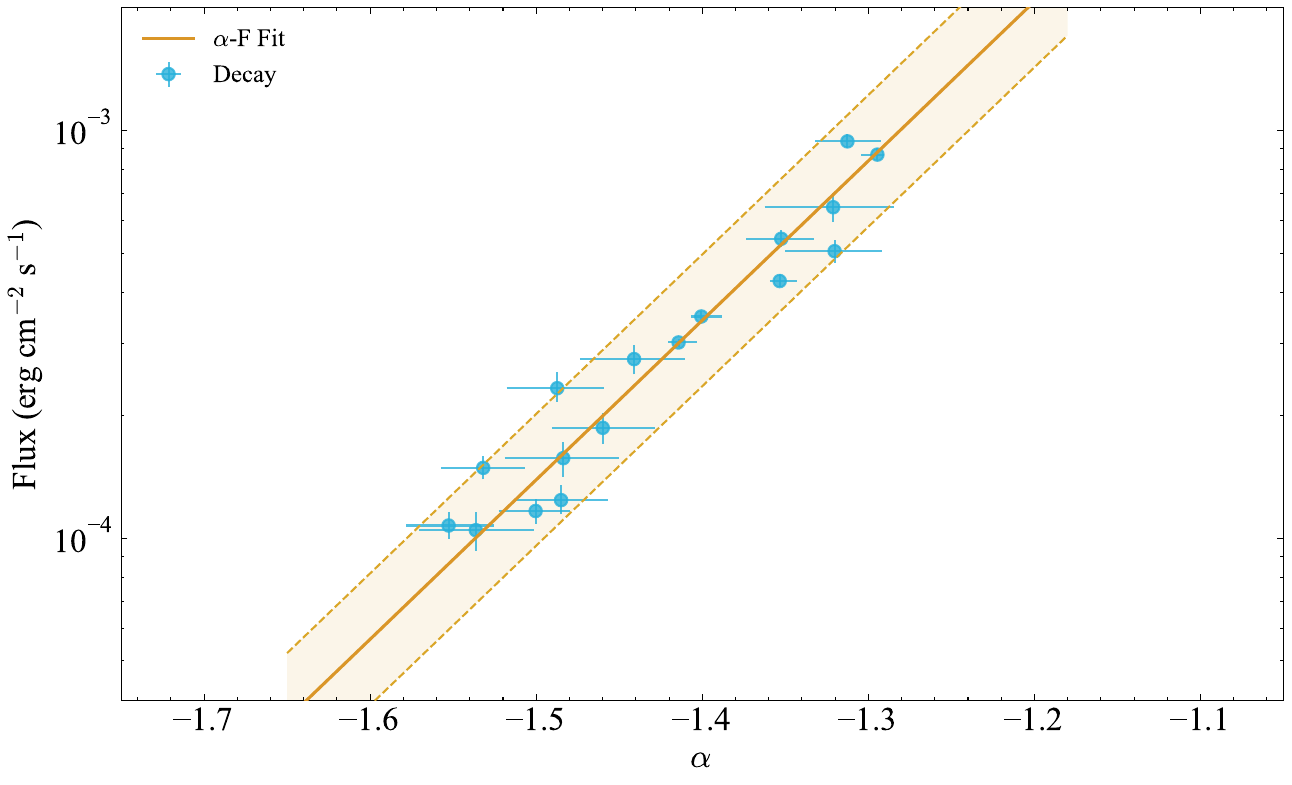}
        \subcaption{}
    \end{minipage}\\
    \begin{minipage}{0.33\textwidth}
        \includegraphics[width = \textwidth]{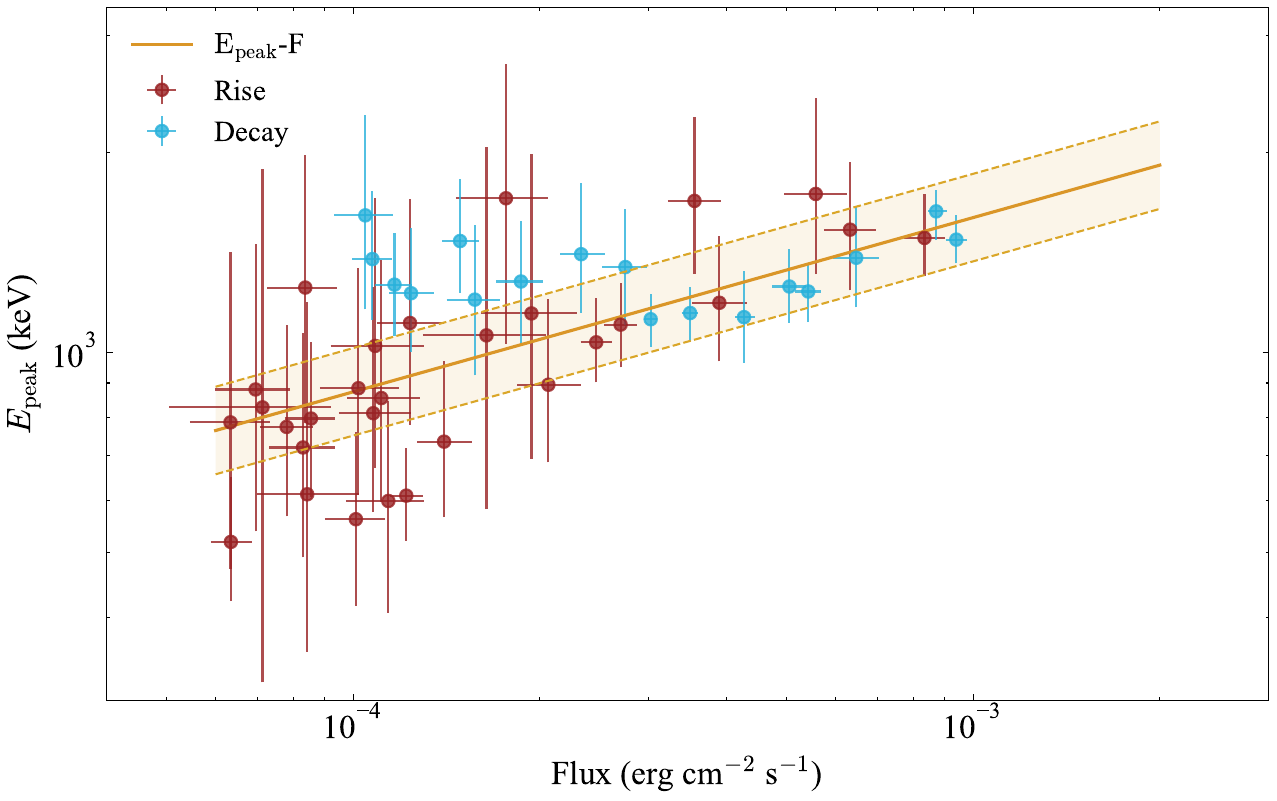}
        \subcaption{}
    \end{minipage}
    \begin{minipage}{0.33\textwidth}
        \includegraphics[width = \textwidth]{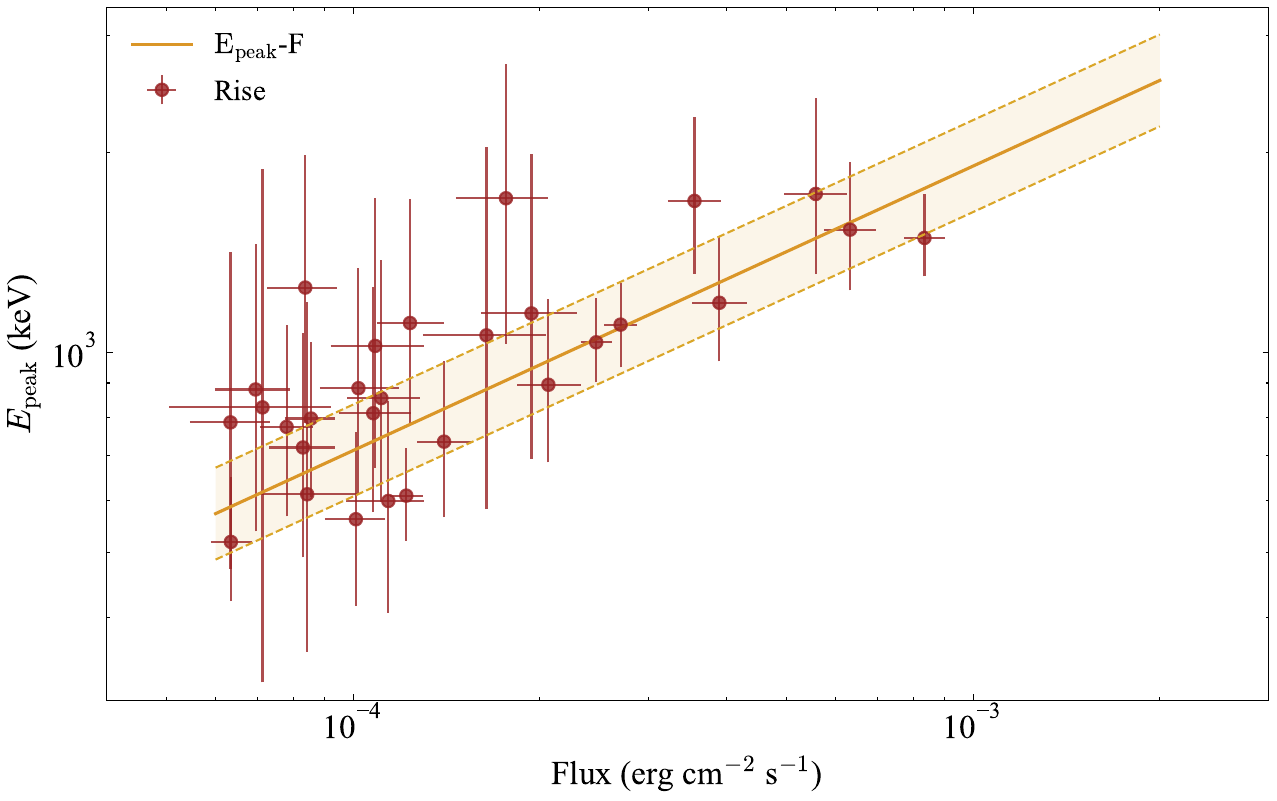}
        \subcaption{}
    \end{minipage}
    \begin{minipage}{0.33\textwidth}
        \includegraphics[width = \textwidth]{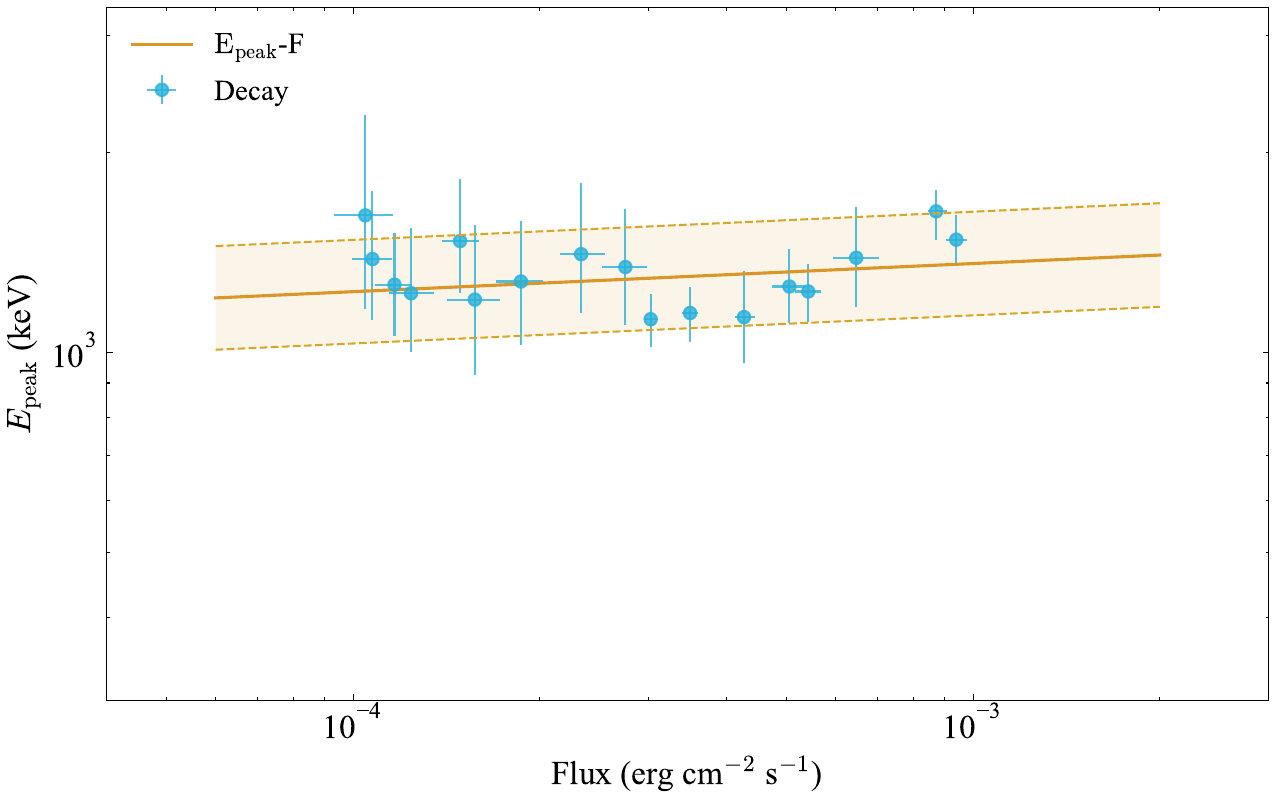}
        \subcaption{}
    \end{minipage}\\
    \begin{minipage}{0.33\textwidth}
        \includegraphics[width = \textwidth]{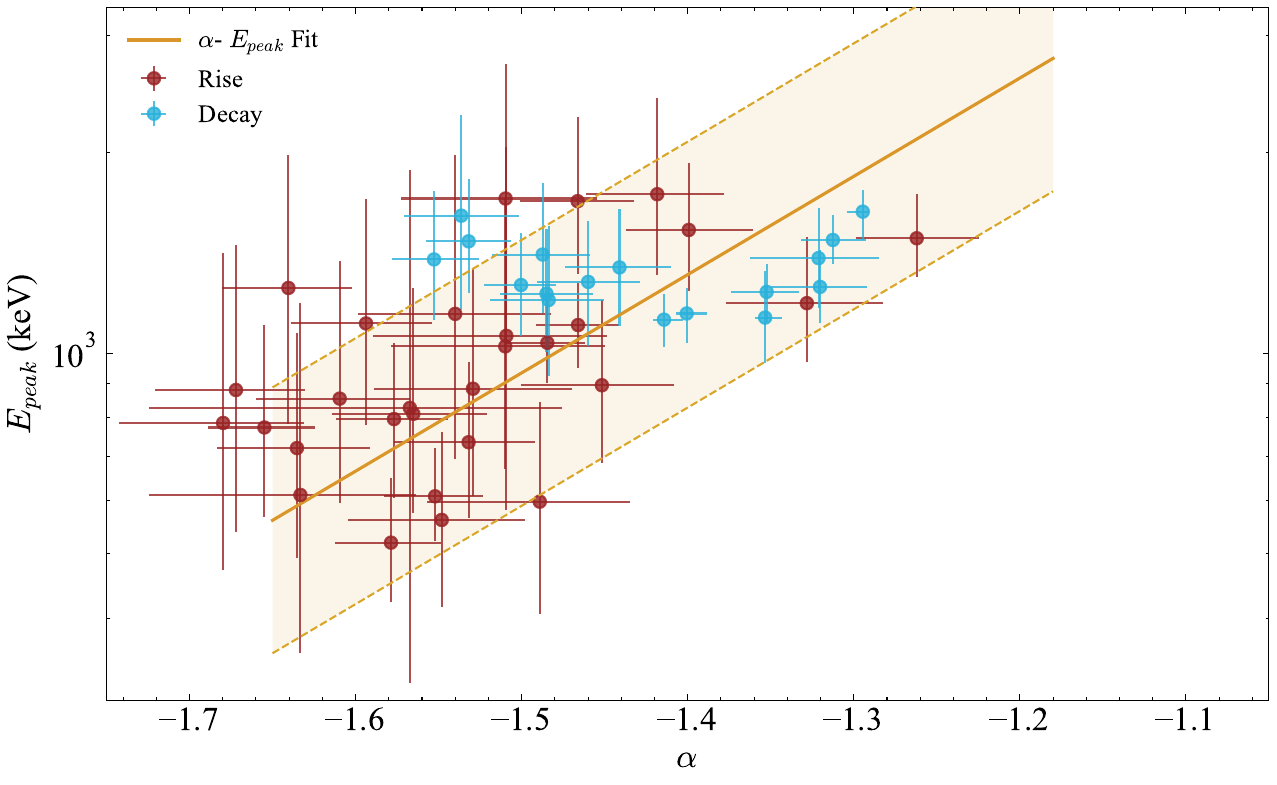}
        \subcaption{}
    \end{minipage}
    \begin{minipage}{0.33\textwidth}
        \includegraphics[width = \textwidth]{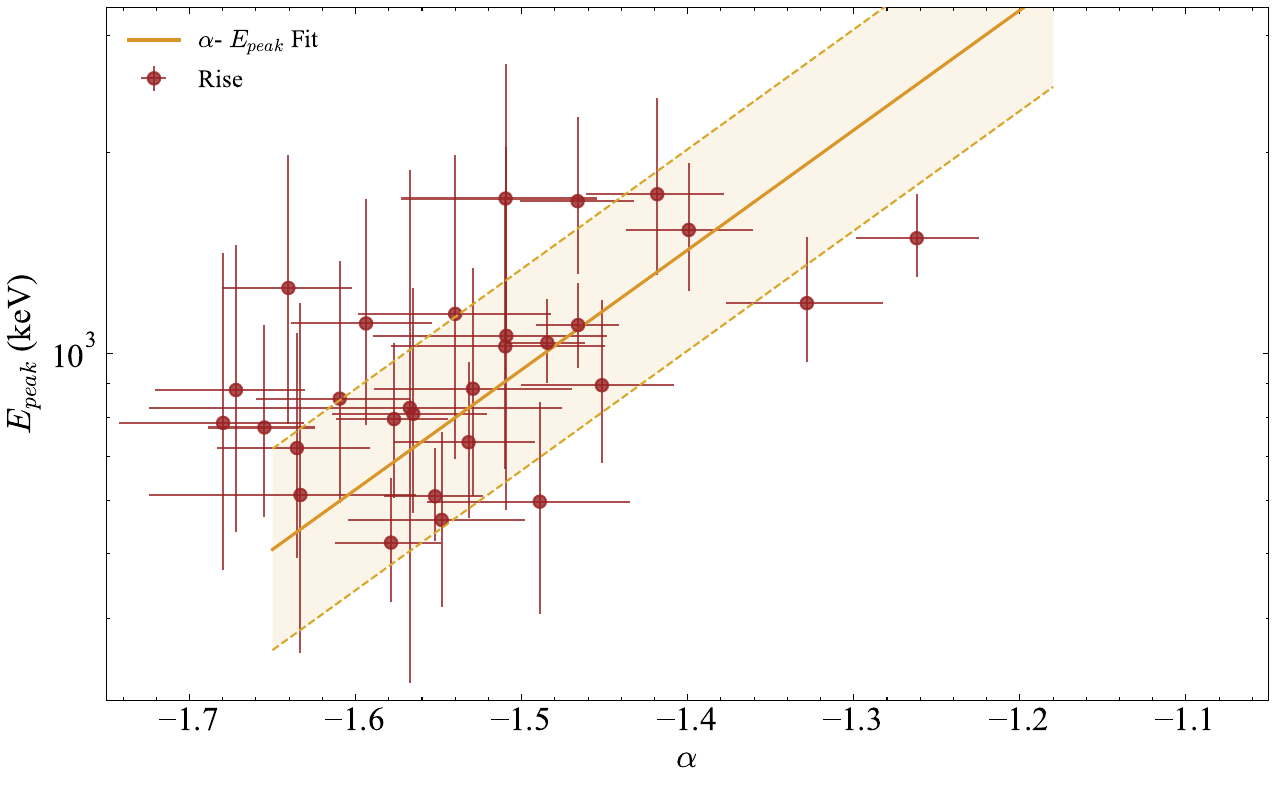}
        \subcaption{}
    \end{minipage}
    \begin{minipage}{0.33\textwidth}
        \includegraphics[width = \textwidth]{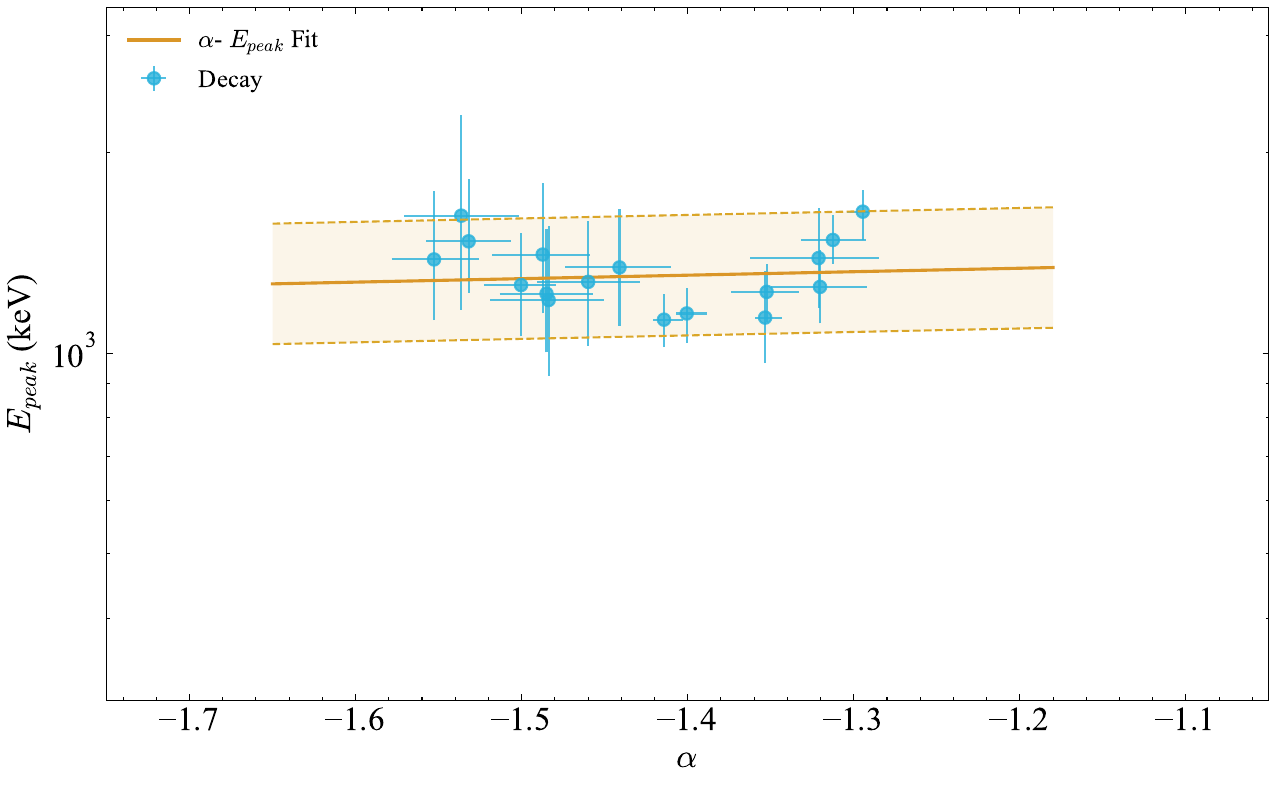}
        \subcaption{}
    \end{minipage}\\
\end{tabular}
  \caption{\label{fig_parameter_correlation}\small The figure shows the relation between spectral parameters. \textbf{Panel (a)–(c)}: The relation between $\alpha$ and $F$ during the overall phase, rise phase, and decay phase are presented. \textbf{Panel (d)–(f)}: They show the relation between $E_{\text{p}}$ and $F$ across the three phases. \textbf{Panel (g)–(i)}: They display the relation between $E_{\text{p}}$ and $\alpha$ for the three phases.}
\end{figure*}

For each time interval, the fitting results of CPL model and Band function are presented separately. The model with a smaller BIC is selected as the spectral parameter result for that time interval. The light curves of the BFL in different energy bands, and the time evolution of the spectral fitting parameters are shown in Figure~\ref{fig1}. The gray dashed lines separate three main pulse components in the light curves and the specific classification criteria have been explained in detail in our previous work.

Figure~\ref{fig_spectrum} displays three spectral fitting results and their corner plots from the total 49 spectral fitting images. Three spectra were successfully fitted using the CPL model. The absence of significant residual evolution structures in the spectral fitting indicates that the fitting results are robust.

The temporal evolution of the spectral fitting results, as shown in Fig.~\ref{fig1}, reveals that both $E_{\text{p}}$ and $\alpha$ exhibit a strong intensity-tracking characteristic. As the flare becomes more intense, both $E_{\text{p}}$ and $\alpha$ significantly harden. This is the first detailed analysis of the high time-resolution spectral evolution in a GRB flare and it offers a possible physical interpretation.
% Interestingly, the ``Double-tracking" behavior has not been identified in previous GRB flares due to their relatively moderate brightness. Therefore this is the first detailed analysis of the ``Double-tracking" feature in a GRB flare and offer a possible physical interpretation.

\subsection{Parameter relation}\label{section2.3}

The relation between spectral parameters plays a crucial role in studying the radiation mechanisms of GRB prompt emission. If the prompt emission and flares of GRBs share a common origin, these parameter relation may also provide insights into the radiation properties of GRB flares. In other words, if the GRB flares exhibit similar spectral parameter relations to the prompt emissions, this would provide strong evidence that they share the same physical origin. These relations of spectral parameters mainly include the relationships between the peak energy ($E_{\text{p}}$), the low-energy spectral index ($\alpha$), and the flux ($F$): $\alpha$-$F$, $E_{\text{p}}$-$F$, and $E_{\text{p}}$-$\alpha$.

\begin{figure*}[htbp]
  \centering
\begin{tabular}{cc}
  \begin{minipage}{0.48\textwidth}
      \includegraphics[width = \textwidth]{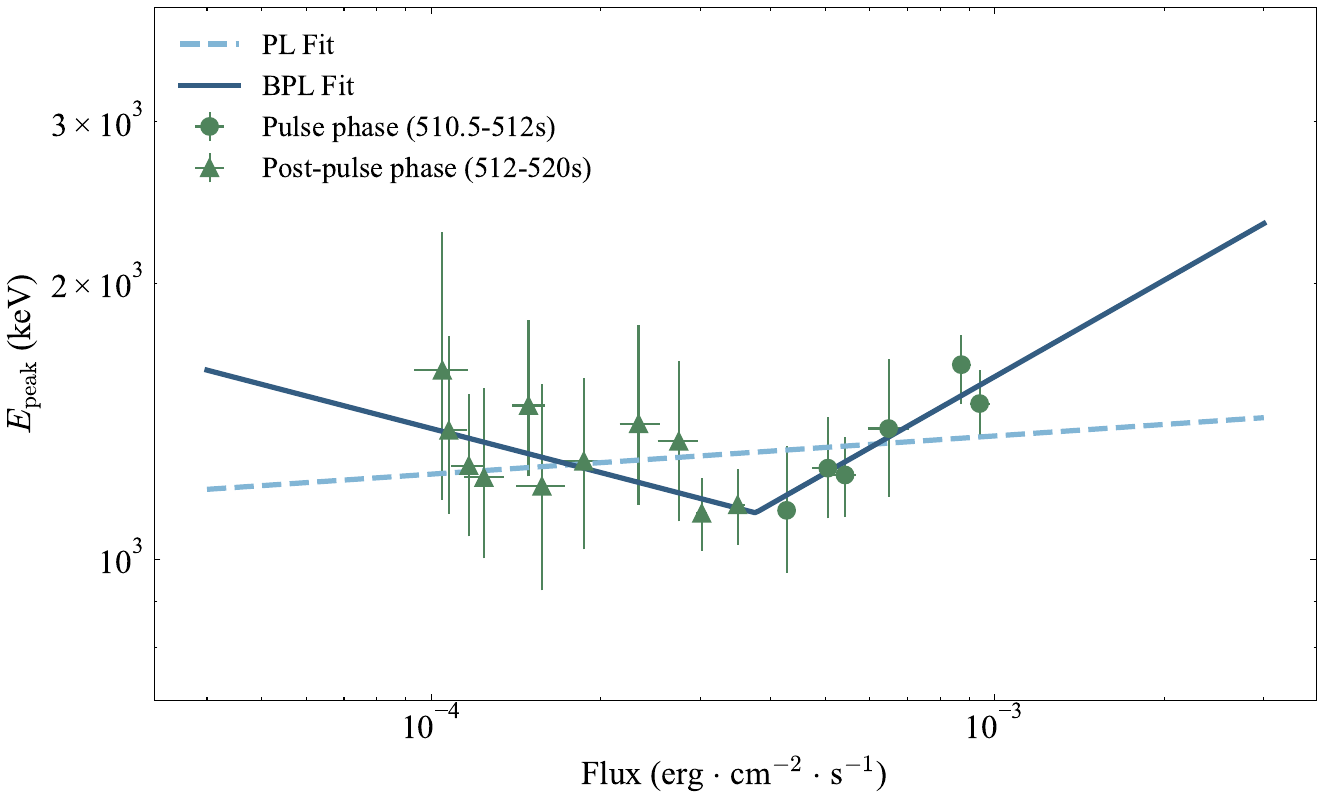}
      \subcaption{}
   \end{minipage}
    \begin{minipage}{0.48\textwidth}
        \includegraphics[width = \textwidth]{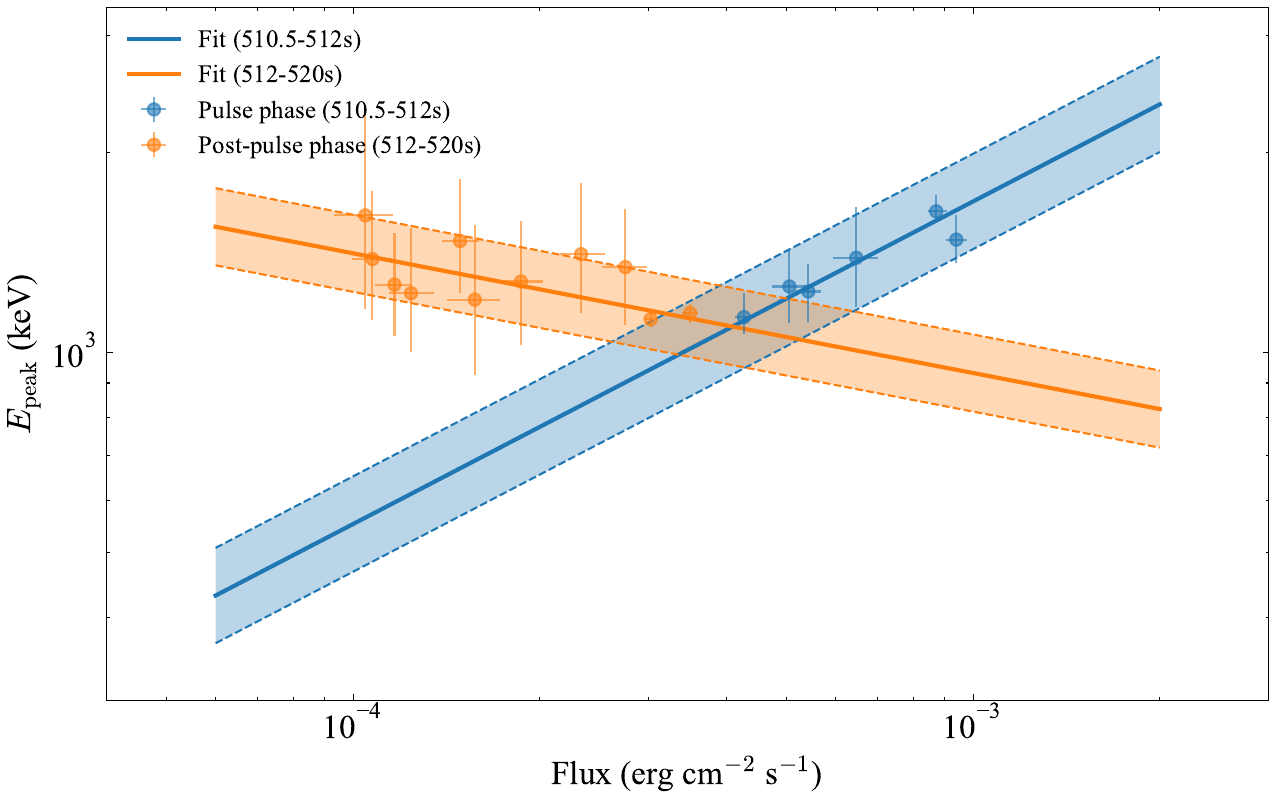}
        \subcaption{}
    \end{minipage}\\
    \begin{minipage}{0.48\textwidth}
        \includegraphics[width = \textwidth]{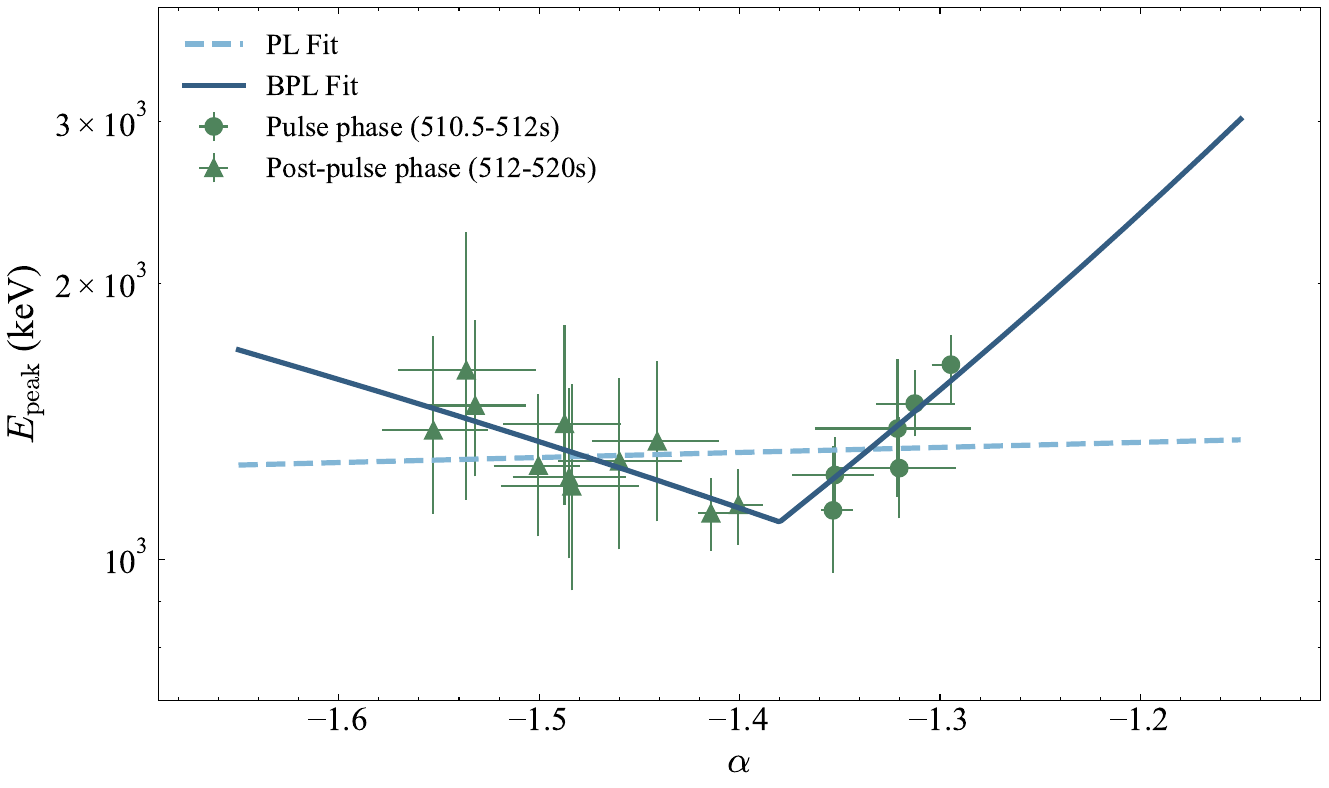}
        \subcaption{}
    \end{minipage}
    \begin{minipage}{0.48\textwidth}
        \includegraphics[width = \textwidth]{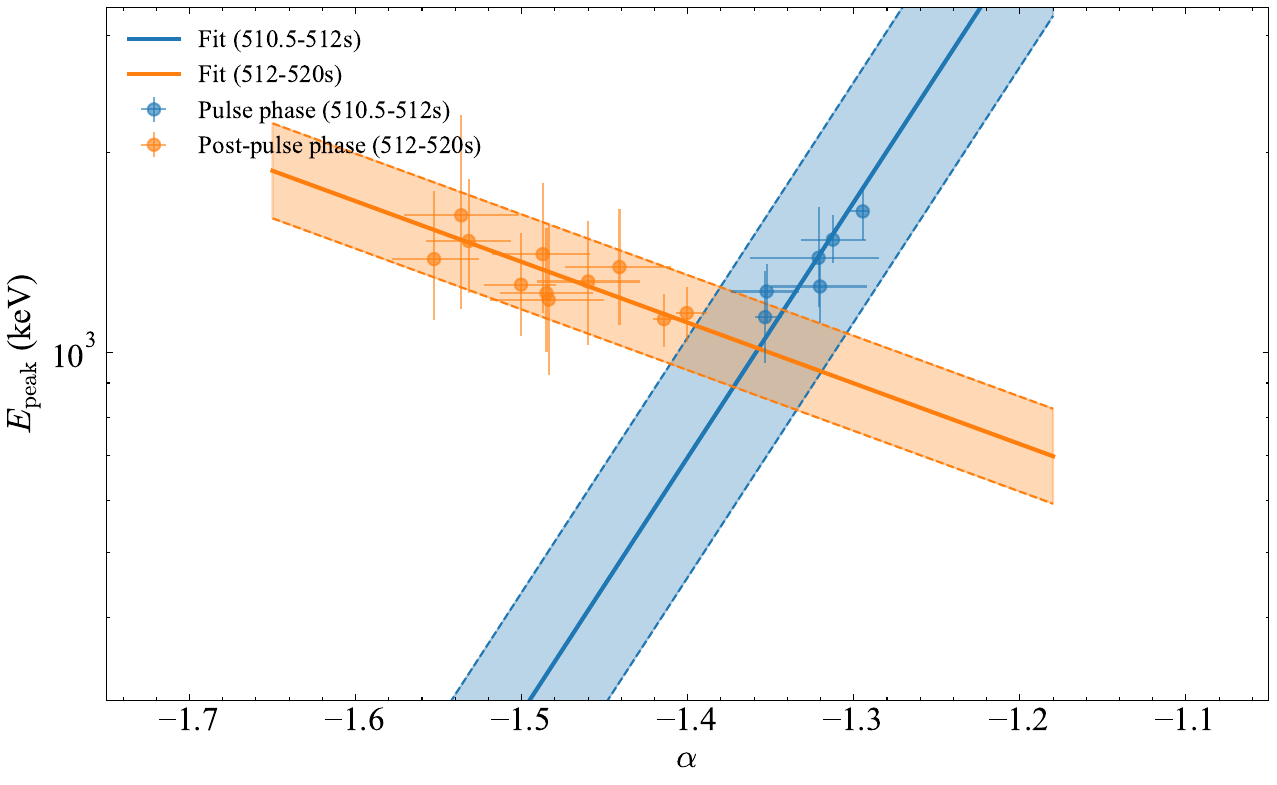}
        \subcaption{}
    \end{minipage}\\

\end{tabular}
  \caption{\label{fig_compare}\small \textbf{Panel (a)-(b)}:Evolution of the $E_{\text{p}} - F$ . \textbf{Panel (a)}: The dark blue curve represents the fitting of the $E_{\text{p}} - F$ relation for the decay phase using the BPL model, while the light blue dashed line represents the fitting using the powerlaw model. The green triangular data points and the green circular data points respectively represent the scatter plots of the $E_{\text{p}} - F$  for the pulse phase and the post-pulse phase. \textbf{Panel (b)}: Blue and orange fitting curves with data points represent the $E_{\text{p}} - F$  fits for the pulse phase and post-pulse phase, respectively. \textbf{Panel (c)-(d)}: Evolution of the $E_{\text{p}} - \alpha$ relation. Legends are similar to those in panel (a) and panel (b).}
\end{figure*}

In our spectral analysis of the BFL, we primarily used the results from the CPL model to investigate the relationships between spectral parameters. The BFL exhibits a light curve structure composed of multiple superimposed pulses, many of which are too narrow to be resolved.  Therefore, we simply divided the flare into the rise phase (from $T_0$+500\,s to $T_0$+510.5\,s) and the decay phase (from $T_0$+510.5\,s to $T_0$+520\,s), using the peak time ($T_0$+510.5\,s) as the boundary. For each pair of spectral parameter relationships, fitting is performed for the rise phase, decay phase, and overall phase (from $T_0$+500\,s to $T_0$+520\,s). To quantitatively evaluate the correlations between different spectral parameters and account for the influence of parameter uncertainties, we performed MCMC simulations to compute the Spearman's rank correlation coefficient $R$ and its uncertainty, along with the corresponding chance probability $p$, for the $\alpha - F$, $E_{\text{p}} - F$ and $E_{\text{p}} - \alpha$ pairs across different time intervals. The resulting correlation measurements are presented in Table~\ref{table_raltion}. In this paper, the MCMC method is mainly employed for fitting the relations. For linear fitting in both log-log and log-linear spaces, the influence of covariance is considered, with specific treatment methods detailed in \cite{B18_230307_nsr}.

\textbf{$\alpha - F$:} The $\alpha$--$F$ relation was fitted in log-linear space using a linear function: $\log F = k \alpha + b$. For the BFL, the best-fit parameter $k$ of overall phase is $3.43 \pm 0.23$, for the rise phase it is $3.49 \pm 0.38$, and for the decay phase it is $3.87 \pm 0.43$. The MCMC fitting results for the $\alpha$-$F$ relationship are shown in Figure~\ref{fig_parameter_correlation} (a)-(c). For the BFL, the rise phase, decay phase, and overall phase exhibit a positive correlation between $\alpha$ and flux. This relationship is commonly observed in previous studies of the relation between spectral parameters during prompt emission and can be well explained by photospheric models or synchrotron models \citep{A22_double_tracking}. All fitting results are summarized in Table~\ref{table_raltion}.

% For comparison, we also applied the fitting form used in \cite{A20_alpha_flux_heat}, $F = N e^{k \alpha}$, to the same data. For the BFL, the best-fit value of k for the overall phase is $7.20 \pm 0.64$, for the rise phase it is $7.60 \pm 0.97$, and for the decay phase it is $7.46 \pm 2.14$. All fitting results are summarized in Table~\ref{table_raltion}.

\textbf{$E_{\text{p}} - F$:} For the relation between $E_{\text{p}}$ and $F$, a linear fit can be performed in the log-log space, which can be described by a simple power-law relation: $E_{\text{p}} = N \cdot F^k$. In previous studies on $E_{\text{p}} - F$ \citep{A12_para_relation_1,B19_Ep_F_previous_study,A13_para_relation_2,A10_EP_F}, the correlation mainly falls into two categories: 1) It includes both power-law segments with positive and negative exponents, exhibiting a break at the peak flux. 2) The relation between $E_{\text{p}}$ and $F$ can be described by a single power-law. In some GRBs, there may also be cases where $E_{\text{p}}$ and $F$ show no significant correlation. For the overall phase of the $E_{\text{p}} - F$ relation of the BFL, the best-fit value of $k$ is $0.27 \pm 0.05$; for the rise phase, it is $0.43 \pm 0.07$; and for the decay phase, it is $0.04 \pm 0.06$. The MCMC fitting results for the $E_{\text{p}} - F$ relation are shown in Figure~\ref{fig_parameter_correlation} (d)-(f). When fitting the rise phase and the overall phase, $E_{\text{p}}$ and $F$ exhibit a clear positive correlation, and the power-law index $k$ obtained from the flare fitting is not significantly different from the previous result for prompt emission $k = 0.55 \pm 0.22$ \citep{A10_EP_F}. However, in the decay phase, the power-law index for the $E_{\text{p}} - F$ relation becomes notably smaller, resulting in a flatter correlation.

During the analysis of the scatter plot of $E_{\text{p}} - F$ in the decay phase, we observed a break in the overall evolutionary trend in the log-log space. This break corresponds precisely to the transition period from the third pulse phase (from $T_0$+511\,s to $T_0$+512\,s) shown in Figure~\ref{fig1} to the decay phase following the pulse phase (from $T_0$+512\,s to $T_0$+520\,s). The decay phase of $E_{\text{p}} - F$ was fitted using a broken power-law (BPL) function, whose mathematical expression is provided in Eq.~\ref{eq_bpl}:

\begin{equation}
E_\text{p}(F) = \begin{cases} 
E_\text{0}(F/F_\text{j})^{-\alpha_1}, & F < F_\text{j} \\ 
E_\text{0}(F/F_\text{j})^{-\alpha_2}, & F > F_\text{j} 
\end{cases}
\label{eq_bpl}
\end{equation}

where $\alpha _1$ and $\alpha _2$ are the temporal slopes, $F_j$ is the break flux, $E_{0}$ is the amplitude.

The Akaike Information Criterion (AIC) is a criterion for model selection, defined as $\text{AIC} = -2 \ln L + 2k$, where $L$ denotes the maximum likelihood value and $k$ is the number of free parameters in the model \citep{1998AIC}. $n$ denotes the number of data points. When $n/k < 40$, we generally adopt the small-sample corrected Akaike Information Criterion ($AIC_c$) as the model selection criterion \citep{2002AICc}. $AIC_c$ is defined as: $AIC_c = -2 \ln L + 2k+2k(k+1)/(n-k-1) = AIC+2k(k+1)/(n-k-1)$ \citep{2002AICc}. We computed both BIC and $AIC_c$ for the power-law (PL) model and BPL model and compare them. It is found that $\Delta \rm BIC = BIC^\text{PL}- BIC^\text{BPL} = 5.09$ and $\Delta \rm AIC_c = AIC_c^\text{PL}- AIC_c^\text{BPL} = 5.92$, which indicates that the BPL model better describes the $E_{\text{p}} - F$ relation during the decay phase compared to the single power-law model. The comparison of the two model fits is presented in Fig.~\ref{fig_compare}(a).

In this paper, we divide the decay phase into two smaller phases: the pulse phase (from $T_0$+510.5\,s to $T_0$+512\,s) and the post-pulse phase (from $T_0$+512\,s to $T_0$+520\,s). The pulse phase is characterized by a prominent pulse, while the post-pulse phase represents the period following the pulse phase. Since the breakpoint of the BPL model occurs precisely at the boundary between the pulse phase and the post-pulse phase, we perform linear fits in log-log space for each phase separately. The best-fit value of $k$ for the pulse phase is $0.42 \pm 0.52$, while for the post-pulse phase, it is $-0.20\pm 0.20$. The MCMC fitting results are shown in Fig.~\ref{fig_compare}(b). Notably, the pulse phase and post-pulse phase exhibit entirely different evolutionary trends. The evolution of $E_{\text{p}} - F$ during the pulse phase shows a positive correlation, consistent with the rise phase and the overall phase trend. However, the $E_{\text{p}} - F$ evolution during the post-pulse phase displays a negative correlation with a lower power-law index, differing from the previous evolution. The physical mechanisms underlying this behavior will be discussed in subsequent sections. The fitting results for the $E_{\text{p}} - F$ relation are summarized in Table~\ref{table_raltion}.

\textbf{$E_{\text{p}}- \alpha:$} The relation between $E_{\text{p}}$ and $\alpha$ can be linearly fitted in the log-linear space, expressed as  $\log E_{\text{p}} = k \alpha + b$.  Previous studies have also provided detailed explanations of $E_{\text{p}} - \alpha$, which exhibits similar characteristics to the $E_{\text{p}} - F$. For the overall phase, the best-fit values of $k$ is $1.78 \pm 0.24$; for the rise phase, it is $1.89 \pm 0.18$; and for the decay phase, it is $0.05 \pm 0.27$. The MCMC fitting plots for $E_{\text{p}} - \alpha$ are shown in Fig.~\ref{fig_parameter_correlation} (g)-(i). $E_{\text{p}} - \alpha$ and $E_{\text{p}} - F$ exhibit strong similarities, showing positive correlations in both the overall phase and rise phase. However, negative correlation emerges during the post-pulse phase, resulting in a relatively ``flat" profile in plot of decay phase.

The scatter plot of $E_{\text{p}} - \alpha$ also exhibits a break during the transition from the pulse phase to the post-pulse phase. We attempted to fit this using the BPL model (log-linear space) and compared it with a simple powerlaw model using the BIC and $AIC_c$. The results show $\Delta \rm BIC = BIC^\text{PL}- BIC^\text{BPL} = 5.42$ and $\Delta \rm AIC_c = AIC_c^\text{PL}- AIC_c^\text{BPL} = 6.89$, indicating that the BPL model better describes the evolution of $E_{\text{p}} - \alpha$ during the decay phase, similar to the $E_{\text{p}} - F$ relation. The comparison plots of the powerlaw model and the BPL model are shown in Fig.~\ref{fig_compare}(c). The best-fit value of $k$ for the pulse phase is $4.50 \pm 1.05$, while for the post-pulse phase, it is $-1.09 \pm 0.73$. The post-pulse phase demonstrates a completely different $E_{\text{p}} - \alpha$ relation compared to the previous phases.

Since both $E_{\text{p}} - F$ and $E_{\text{p}} - \alpha$ revealed that the post-pulse phase has different relation compared to the earlier phases, we also examined the $\alpha - F$ relation in both the pulse phase and the post-pulse phase. The best-fit value of $k$ for the $\alpha - F$ in the pulse phase is $6.57 \pm 1.23$, while for the post-pulse phase, it is $3.63 \pm 1.41$. This is consistent with the evolutionary trends observed in the overall, rise phase, and decay phase, where the $\alpha - F$ relationship exhibits a clear ``flux-tracking" property across all phases. Due to this characteristic, the phenomenon that $E_{\text{p}} - F$ and $E_{\text{p}} - \alpha$ share the same evolutionary pattern becomes more natural. The fitting results are summarized in Table~\ref{table_raltion}.

$E_\text{p}$ and $\alpha$ both exhibit an overall ``flux-tracking" behavior. During the rise phase as well as in the overall fit, $\alpha$ shows a positive correlation with $F$, $E_\text{p}$ with $F$, and $E_\text{p}$ with $\alpha$. In the decay phase, $\alpha$ remains positively correlated with $F$, whereas the relationship curves between $E_\text{p}$ and $F$ and between $E_\text{p}$ and $\alpha$ become flatter, with their power-law indices decreasing significantly. Compared to a simple power-law function, a broken power-law function provides a better fit to the correlations of $E_\text{p}$-$F$ and $E_\text{p}$-$\alpha$. By dividing the decay phase at $T_0$+512\,s into the pulse phase and the post-pulse phase, we find that $E_\text{p}$ remains positively correlated with both $F$ and $\alpha$ during the pulse phase, while these correlations turn negative in the post-pulse phase. In Section~\ref{section3}, we will provide possible physical interpretation for the aforementioned phenomenon.

Although ``Double-tracking" is commonly observed during the prompt emission \citep{A22_double_tracking,A21_broken_alpha_flux}, it had not previously been detected in GRB flares, mainly due to limitations in flare brightness and instrumental sensitivity. Thanks to the exceptional brightness of the BFL in GRB 221009A, we have, for the first time, observed the ``Double-tracking" phenomenon in a GRB flare. This provides another strong piece of evidence supporting the hypothesis that GRB prompt emission and GRB flare share the same physical origin. Furthermore, it offers a new ``probe" for studying the radiation mechanisms of GRB flares and GRB prompt emission.

\section{Physical Interpretation} \label{section3}

When performing the spectral fitting on the BFL,, the non-thermal spectrum can be applied to all time intervals, and the low-energy spectral index $\alpha$ in all time intervals does not exceed the ``death line" ($\alpha = -2/3$) \citep{C1_1994_alpha_2/3,C2_1998_2/3_2}. Therefore, in the context of the BFL, we primarily employ synchrotron-dominated radiation models to explain the relations between the spectral parameters of this flare.

In Section~\ref{section3.1} and Section~\ref{section3.2}, we discuss the explanation of the ``flux-tracking" phenomenon of $E_{\text{p}} - F$ and $\alpha - F$ within the synchrotron radiation framework. For the $E_{\text{p}}$, our explanation primarily focuses on this phenomenon during the rise phase of the flare. In Section~\ref{section3.3}, we provide a possible explanation for why the decay phase exhibits different evolution compared to the rise phase.

\subsection{Radiation physics of $E_{\text{p}} - F$ correlation } \label{section3.1}

The previous study of GRB 221009A suggests that it had a highly magnetized, Poynting-flux-dominated jet \citep{C9_2023_flare_explain}. In this work, we consider using the magnetic dissipation model within the synchrotron radiation framework to explain the relations between spectral parameters. The radiation radius can be expressed as: $R \sim \Gamma^2 c t_{\text{pulse}}$, where $t_{\text{pulse}}$ is the duration of the pulse, according to the Internal-Collision-induced Magnetic Reconnection and Turbulence (ICMART) model \citep{C3_2011_ICMART}. The BFL exhibits a multi-peaked temporal structure. In this paper, we treat the entire flare as a single large pulse, and the rapid variability time scale within this large pulse are thought to be possibly caused by localized magnetic reconnection events within the jet \citep{c4_mini_reconnection}.

In the framework of synchrotron radiation, $E_{\text{p}} \propto L^{1/2} \gamma_{e}^2 R^{-1}/(1+z)$, where $L$ represents the luminosity of the jet, $\gamma_{\text{e}}^2$ is the typical Lorentz factor of the electrons in the radiation region, $R$ denotes the radiation radius, and $z$ is the redshift \citep{C5_2002_Ep_L,A22_double_tracking}. When the parameters in this equation are appropriately adjusted, a scenario arises where  $E_{\text{p}} \propto L^{1/2}$. Since $L$ and $F$ are directly proportional, this naturally leads to the ``flux-tracking" phenomenon in $E_{\text{p}}$.

In Section~\ref{section2.3}, the power-law indices obtained from the $E_\text{p}-F$ fitting for both the rise phase ($k=0.43 \pm 0.07$) and the pulse phase ($k=0.42 \pm 0.52$) are approximately 0.5, indicating that the fitting results are consistent with the theoretical interpretation. 
% In Section~\ref{section3.3}, we will address the anomalous inverse correlation observed between $E_{\text{p}}$ and $F$ during the post-pulse phase of the flare, and propose potential explanations for this behavior.

\subsection{Radiation physics of $\alpha-F$ correlation }\label{section3.2}

As the jet expands, the emission radius gradually increases as the emission region moves further away from the central engine. During this process, the magnetic field decays with the relation $B \propto R^{-b}$. The decay index $b$ of the magnetic field, together with the electron injection rate, jointly regulates spectral index $\alpha$ \citep{A19_decayB_alpha_flux}. This may lead to a transition between fast cooling and slow cooling in the radiation region, which could result in the ``flux-tracking" behavior of $\alpha$.

Furthermore, previous research has also indicated that the ``flux-tracking" behavior of $\alpha$ could be attributed to synchrotron self-Compton scattering (SSC) \citep{c7_2009_ssc_alpha_2,C6_2011_SSC_ALPHA,c8_2018_ssc_alpha,A22_double_tracking}. As mentioned in Section~\ref{section3.1}, $E_{\text{p}}$ depends on the luminosity $L$, the Lorentz factor $\gamma_{\text{e}} ^2$ of the electrons, and the emission radius $R$. During the rise phase, $E_{\text{p}}$ increases with time, while $R$ also expands. Since $E_{\text{p}} \propto R^{-1}$, to ensure that both $E_{\text{p}}$ and $R$ increase simultaneously, $\gamma_{\text{e}}$ must increase, resulting in hardening of $\alpha$. The increase in $\gamma_{\text{e}}$ would then enhance the SSC process \citep{c8_2018_ssc_alpha}. The enhancement of SSC may result in the emission energy density becoming dominant, which naturally leads to an increase in the emission flux. Under the interaction of these factors, during the rise phase, as the flux increases, $\alpha$ hardens. During the decay phase, it is possible that the increase in $\gamma_{\text{e}}$ may diminish or gradually weaken, resulting in a decrease in flux and a softening of $\alpha$. This could also be one of the reasons for the ``flux-tracking" behavior in $\alpha$.

\subsection{Parameter evolution in the decay phase } \label{section3.3}

The decay phase of the BFL exhibits a distinctly different spectral evolution pattern in the $E_{\text{p}} - F$ compared to the rise phase, with an overall ``flatter" power-law index. When fitting the $E_{\text{p}} - F$ relation for the decay phase, the BPL model provides a better fit than a simple power-law model.  

Interestingly, when we separately perform linear fits in log–log space to the two phases of the decay phase, which are the pulse phase and the post-pulse phase, the pulse phase exhibits a ``flux-tracking" behavior in $E_{\text{p}}$, similar to the rise phase (Fig.~\ref{fig_compare}). However,the post-pulse phase displays negative correlation between $E_{\text{p}}$ and $F$ (Fig.~\ref{fig_compare}). As shown in Fig.~\ref{fig1} and Fig.~\ref{fig_compare}, during the post-phase phase, $E_{\text{p}}$ tends to harden as the $F$ softens.

In previous studies on the GRB 221009A flare, we fitted the light curve of the GRB 221009A flare, revealing that GRB 221009A flare is composed of multiple flares \citep{09A_flare}. The brightest one among them is the subject of this work. Through fitting, we discover that the subsequent flare begins to erupt even before the BFL finished. The observed decay phase of the BFL is likely the result of the combined effects of the BFL decay phase and the rise phase of the subsequent flare.

This provides a new perspective for understanding the anomalous evolution of the $E_\text{p}$-$F$ relation during the decay phase. In the pulse phase, the emission remains dominated by the BFL component, resulting in the same spectral parameter evolution as in the rise phase. However, during the post-pulse phase, the contribution from the subsequent flare gradually increases. As shown in the zoomed-in panel of Figure~\ref{fig1}(c), during the post-pulse phase, the subsequent flare following the BFL has already begun its eruption. We argue that this likely causes the gradual hardening of $E_{\text{p}}$ while the flux decreases in the post-pulse phase. Such anomalous tracking evolution is essentially a combined outcome of the concurrent evolution of the BFL and the subsequent flare.

%The combined effects of the two flares may influence the behavior of $E_{\text{p}}$, causing it to harden even as the flux softens. This leads to different spectral parameter evolution trends between the pulse phase and the post-phase phase. 

Therefore, when fitting the $E_{\text{p}} - F$ during the decay phase, BPL model provides a better description, and when using a simple power-law function to fit the decay phase, the power-law index is smaller, resulting in a ``flatter" curve. Since the $\alpha-F$ relation shows positive correlation across all phases, it is relatively straightforward to understand that the evolution of $E_{\text{p}} - \alpha$ is similar to that of $E_{\text{p}} - F$.

\subsection{Discussion}\label{section4}

% In this study, we find that GRB flares and prompt emission exhibit similar evolutionary patterns in their spectral parameters, which provides another evidence supporting the notion that prompt emission and flare share a similar physical origin. In this paper, we focuses on constructing a plausible physical model to explain such phenomenon.

In previous research on the physical explanation of the ``$\alpha$-intensity" (``flux-tracking" in $\alpha$), the entropy-dominated photospheric heating model was considered more natural due to the presence of thermal components in some GRBs \citep{A20_alpha_flux_heat}. In this model, the co-evolution of $\alpha$ and $F$ is driven by the temporal evolution of the dimensionless jet entropy $\eta$, which reflects the relationship between the photospheric radius and the saturation radius. 
%When $\eta$ is high, the photosphere is close to the saturation radius, resulting in high flux and a hard spectrum. Conversely, when $\eta$ is low, the photosphere moves away from the saturation radius, leading to low flux and a soft spectrum. 
The $\eta$-driven mechanism naturally explains the observed flux-tracking behavior of $\alpha$. Regarding the empirically found relation $F \propto e^{k \alpha}$, with a median $k \approx 3$, \cite{A20_alpha_flux_heat} provided theoretical support by comparing energy losses between the saturation radius and the Wien radius. Since the flux is governed by adiabatic cooling following $F \propto r^{-2/3}$, under typical optically thick conditions (optical depth $\tau \sim 100$), the decrease in luminosity during spectral evolution from hard to soft corresponds to a variation in $\alpha$ that yields $k \approx \frac{2}{3}ln(100) \approx 3.07$, which aligns closely with observations. It is worth noting that the sample utilized for the $\alpha-F$ correlation study predominantly consists of GRBs , which exhibit $\alpha$ exceeding the ``death line'' during certain intervals \citep{A20_alpha_flux_heat}.

For the BFL, the absence of significant thermal components does not completely rule out a photospheric origin. Previous studies have demonstrated that photospheric models \citep{C105_2005_Rees_photo,C108_2006_pe_photo,C102_2008_Giannios_photo,C103_2011_Beloborodov_photo,C109_Ryde_2011_photo,C106_2013_Lundman_photo,C101_2014_Deng_Ep_photo,C107_2018_meng_photo} can also produce non-thermal spectra. If the BFL originates from a photospheric model, the observed $E_{\text{p}} - F$ tracking could still be explained theoretically. Within the photospheric framework, $E_{\text{p}}$ is correlated with temperature. A flux-tracking pattern can emerge if the dimensionless entropy $\eta$ exhibits a certain power-law dependence on the wind luminosity $L_{\rm w}$ \citep{d1_2012_Ep_heat,C101_2014_Deng_Ep_photo}.

However, explaining the simultaneous tracking of $\alpha - F$ presents substantial difficulties. Photospheric models typically predict a harder low-energy spectral index ($\alpha \sim +0.5$) \citep{C101_2014_Deng_Ep_photo}. In contrast, $\alpha$ never exceeds $-1$ throughout the BFL in our observations. Furthermore, explaining the smooth tracking behavior of such a soft, non-thermal $\alpha$ would require highly contrived assumptions about the jet structure and its temporal evolution \citep{C106_2013_Lundman_photo,A22_double_tracking}. Only under such fine-tuned conditions could the photospheric model potentially account for this phenomenon. Therefore, we intend to conclude that the consistently soft $\alpha$ of the BFL, together with its smooth "Double-tracking" behavior, disfavors a photospheric interpretation.

% However, since the BFL lacks thermal component, and 
Previous studies of GRB 221009A flare have demonstrated the viability of a magnetic dissipation model \citep{C9_2023_flare_explain}, we adopt a magnetic dissipation model within synchrotron radiation framework to explain the observed spectral evolution.

During the decay phase, the power-law index of $E_{\text{p}} - F$ becomes smaller and approaches 0, and BPL model provides a better fit than power-law model, which suggests the possible existence of two flare components. The different evolution of spectral parameters during the pulse phase and post-pulse phases further support this physical explanation.

During the decay phase, the unusual evolutionary behavior of $E_{\text{p}}-F$ and $E_{\text{p}}-\alpha$ suggests that the superposition of multiple pulses likely modulates the correlation evolution among spectral parameters. This inter‑pulse interference can cause an intrinsic power‑law correlation to appear observationally as a BPL structure. We argue that, for a GRB with complex multi‑pulse morphology such as GRB 230307A, the anomalous BPL evolution seen in its $\alpha-F$ relation may also be related to this effect \citep{A21_broken_alpha_flux}. This inference is further supported by a reverse test: in a typical single pulse burst like GRB 131231A \citep{A22_double_tracking}, where the evolutionary track is not contaminated by subsequent or concurrent pulses, no such distortion of the correlation is observed.

\begin{deluxetable*}{cccccccccccc}
%\tabletypesize{\scriptsize}
\setlength{\tabcolsep}{3pt}
\tablewidth{3pt}
\tablecaption{Fitting results of the parameter relation for the BFL \label{table_raltion}}
\tablehead{
 & & \multicolumn{2}{c}{$\text{Power-law}^\textbf{a}$} & \multicolumn{4}{c}{\text{Broken Power-law}} & \colhead{} & \colhead{} & \colhead{} & \colhead{}\\
\cline{3-8}
\colhead{Relation} & \colhead{Phase} & \colhead{k} & \colhead{b} & \colhead{$\alpha_1$} & \colhead{$\alpha_2$} & \colhead{Break} & \colhead{$E_{\text{0}}$} & \colhead{${\Delta BIC}^\textbf{b}$} & \colhead{$\Delta AIC_c$} & \colhead{R} & \colhead{p}\\
 \colhead{} & \colhead{} & \colhead{} & \colhead{} & \colhead{} & \colhead{} & \colhead{} & \colhead{(keV)} & \colhead{} & \colhead{} & \colhead{} & \colhead{}
}
\startdata
\multirow{5}{*}{\textbf{$\alpha - F$}} & Overall & $3.43 \pm 0.23$ & $1.38 \pm 0.33$ & -- & -- & -- & -- & -- & -- & $0.81 \pm 0.05$ & $<10^{-3}$\\
  & Rise & $3.49 \pm 0.38$ & $1.54 \pm 0.59$ & -- & -- & -- & -- & -- & -- & $0.69 \pm 0.09$ & $<10^{-3}$\\
  & Decay & $3.87 \pm 0.43$ & $2.04 \pm 0.62$ & -- & -- & -- & -- & -- & -- & $0.90 \pm 0.03$ & $<10^{-3}$\\
  & Pulse & $6.57 \pm 1.23$ & $5.53 \pm 1.64$ & -- & -- & -- & -- & -- & -- & $0.52 \pm 0.25$ & 0.072\\
  & Post-Pulse & $3.63 \pm 1.41$ & $2.43 \pm 1.82$ & -- & -- & -- & -- & -- & -- & $0.75 \pm 0.11$ & $<10^{-3}$\\
\hline
\multirow{5}{*}{\textbf{${E_{\text{p}} - F}^\textbf{c}$}} & Overall & $ 0.27 \pm 0.05$ & $3.92 \pm 0.16$ & -- & -- & -- & -- & -- & -- & $0.56 \pm 0.08$ & $<10^{-3}$\\
  & Rise & $0.43 \pm 0.07$ & $4.54 \pm 0.27$ & -- & -- & -- & -- & -- & -- & $0.62 \pm 0.09$ & $<10^{-3}$\\
  & Decay & $0.04 \pm 0.06$ & $3.28 \pm 0.22$ & $0.18 \pm 0.11$ & $-0.38 \pm 0.13$ & $(3.74 \pm 0.49) \times 10^{-4}$ & $1108.40 \pm 75.33$ & 5.09 & 5.92 & $0.07 \pm 0.21$ & 0.863\\
  & Pulse & $0.42 \pm 0.52$ & $4.36 \pm 1.58$ & -- & -- & -- & -- & -- & -- & $0.77 \pm 0.11$ & 0.019\\
  & Post-Pulse & $-0.20 \pm 0.20$ & $2.36 \pm 0.75 $ & -- & -- & -- & -- & -- & -- & $-0.31 \pm 0.25$ & 0.077\\
\hline
\multirow{5}{*}{\textbf{${E_{\text{p}} - \alpha}^\textbf{d}$}} & Overall & $1.78 \pm 0.24$ & $5.65 \pm 0.36$ & -- & -- & -- & -- & -- & -- & $0.45 \pm 0.12$ & $<10^{-3}$\\
  & Rise & $1.89 \pm 0.18$ & $5.98 \pm 0.27$ & -- & -- & -- & -- & -- & -- & $0.37 \pm 0.15$ & 0.002\\
  & Decay & $0.05 \pm 0.27$ & $3.27 \pm 0.38$ & $-2.64 \pm 1.73$ & $5.04 \pm 2.51$ & $-1.38 \pm 0.02$ & $1076.08 \pm 95.81$ & 5.42 & 6.89 & $-0.08 \pm 0.21$ & 0.793\\
  & Pulse & $4.50 \pm 1.05$ & $9.10 \pm 1.41$ & -- & -- & -- & -- & -- & -- & $0.57 \pm 0.33$ & 0.005\\
  & Post-Pulse & $-1.09 \pm 0.73$ & $1.48 \pm 1.09$ & -- & -- & -- & -- & -- & -- & $-0.58 \pm 0.21$ & 0.007\\
\hline
\enddata
\tablecomments{
\textbf{a:} In this paper, the power-law relation is obtained through linear fitting in log-log space or log-linear space. For $\alpha - F$: $\log F = k \alpha + b$; for $E_{\text{p}} - F$: $\log E_{\text{p}} = k\log F + b$; for $E_{\text{p}}- \alpha$: $\log E_{\text{p}} = k \alpha + b$. Through fitting, we can derive $k$ and $b$. \textbf{b:} $\Delta BIC = BIC^\text{PL}- BIC^\text{BPL}$. $\Delta AIC_c$ is calculated in the same manner. \textbf{c:} The unit of $E_{\text{p}}$ is keV, and the unit of $F$(flux) is $\text{erg} \cdot \text{cm}^{-2} \cdot \text{s}^{-1}$. In $E_{\text{p}} - F$, the break is $F_\text{j}$, which is the break point of flux. \textbf{d:} In $E_{\text{p}} - \alpha$, the break is $\alpha_\text{j}$, which is the break point of low enegry index $\alpha$.
}
\end{deluxetable*}

\section{Summary}\label{section4}

In this paper, we report the evolution of $E_{\text{p}}$ and $\alpha$ in the BFL observed in GRB 221009A. We perform, for the first time, a high-resolution temporal spectral analysis of a GRB flare, which reveals the evolution of spectral parameters within a single GRB flare. The fitting of the spectral parameter correlations is primarily analyzed over three time intervals: the rise phase (from $T_0$+500\,s to $T_0$+510.5\,s), the decay phase (from $T_0$+510.5\,s to $T_0$+520\,s), and the overall phase(from $T_0$+500\,s to $T_0$+520\,s). 

Strong positive correlations are found between $E_{\text{p}} - F$, $\alpha - F$, and $E_{\text{p}} - \alpha$ in both the rise phase and the overall phase. This is the first observation of the ``Double-tracking" behavior in a GRB flare, where both $E_{\text{p}}$ and $\alpha$ exhibit ``flux-tracking" behavior. This provides strong evidence supporting the notion that the prompt emission and GRB flares share the same physical origin. We employ a magnetic dissipation model within the synchrotron radiation framework to offer a possible physical explanation for this ``flux-tracking" behavior. The $E_{\text{p}} - F$ relation can be explained by the relation $E_{\text{p}} \propto L^{1/2} \gamma_{e}^2 R^{-1} /(1+z)$. According to the synchrotron radiation theory, a correlation between the $E_{\text{p}}$ and $F$ is expected, typically in the form of $E_{\text{p}} \propto L^{1/2}$. Our fitting to the $E_{\text{p}} - F$ relation during the rise phase and pulse phase yields a power-law index around 0.5, which is consistent with the theoretical prediction.
The $\alpha - F$ relation may result from the combined effects of the magnetic field decay index $b$ and the electron injection rate, or possibly be due to synchrotron self-Compton scattering.

In the decay phase, the power-law indices of $E_{\text{p}} - F$ and $E_{\text{p}} - \alpha$ significantly decrease, and the correlation plots between spectral parameters become ``flatter." This phenomenon suggests that, in addition to the BFL itself, there may be other components present during the decay phase. In this paper, the decay phase is divided into two phases: the pulse phase (from $T_0$+510.5\,s to $T_0$+512\,s) and the post-pulse phase (from $T_0$+512\,s to $T_0$+520\,s). During the pulse phase, $E_{\text{p}} - F$ and $E_{\text{p}} - \alpha$ exhibit a positive correlation similar to that in the rise phase. In contrast, the post-pulse phase exhibits negative correlations with significantly strongly negative power-law indices which is different from the rise phase. Previous studies have found that while the BFL is still in the decay phase, the subsequent flare has already begun to erupt \citep{09A_flare}. Therefore, we believe that the observed evolution of $E_{\text{p}} - F$ and $E_{\text{p}} - \alpha$ may result from the combined effects of both the BFL and the subsequent flare.

This study of spectral parameter correlations in the BFL of GRB 221009A provides a new ``probe" for understanding the physical picture and radiation mechanisms of GRB flares and prompt emission, offering fresh insights into the physical processes underlying such events.

\begin{acknowledgments}
This work is supported by the National Natural Science Foundation of China (Grant Nos. 
12494572, 12494570%SVOM, Xiongshaolin
,
12273042% GECAM, Xiongshaolin
), the Strategic Priority Research Program of the Chinese Academy of Sciences (Grant No. XDB0550300%HXMT GECAM, Xiongshaolin(taolian)
) and 
the China's Space Origins Exploration Program. %% eXTP, Xiongshaolin
%the National Key R\&D Program of China (2021YFA0718500)
The GECAM (Huairou-1) mission is supported by the Strategic Priority Research Program on Space Science (Grant No. XDA15360000) of Chinese Academy of Sciences. We appreciate the GECAM team and the SATech-01 satellite team who made this observation possible.
\end{acknowledgments}

%\begin{contribution}

%All authors contributed equally to the Terra Mater collaboration.

%\end{contribution}

%\clearpage

\bibliography{sample701}{}
\bibliographystyle{aasjournalv7}

\appendix
\renewcommand{\thefigure}{\thesection.\arabic{figure}} 
\renewcommand{\thetable}{\thesection.\arabic{table}}
\setcounter{figure}{0} 
\setcounter{table}{0}

\section{Spectra fitting results}

\begin{longtable}{cccccccccc}
\scriptsize
\renewcommand{\arraystretch}{0.3}
\setlength{\tabcolsep}{2pt} \\
\caption{Time resolved spectra fitting result of the Brightest flare}\\
\label{table_spectrum}\\
\toprule[1.5pt]
\midrule[0.8pt]
    {Start Time} & {End Time} & {model} & {$\alpha$} & {$\beta$} & {$E_{\text{p}}$} & {Flux} & {cstat/dof}& {BIC} & {Best model} \\
    {(s)} & {(s)} & {} & {} & {} & {(keV)} & {$10^{-5} \text{erg} \cdot \text{cm}^{-2} \cdot \text{s}^{-1} $} & {} & {} & {}\\
\midrule[0.8pt]
\endfirsthead
\midrule[0.8pt]
\caption{continue}\\
\toprule[1.5pt]
\midrule[0.8pt]
    {Start Time} & {End Time} & {model} & {$\alpha$} & {$\beta$} & {$E_{\text{p}}$} & {Flux} & {cstat/dof}& {BIC} & {Best model} \\
    {(s)} & {(s)} & {} & {} & {} & {(keV)} & {$10^{-5} \text{erg} \cdot \text{cm}^{-2} \cdot \text{s}^{-1} $} & {} & {} & {}\\
\midrule[0.8pt]
\endhead
\midrule[0.8pt]
\endfoot
\caption{continue}\\
\toprule[1.5pt]
\midrule[0.8pt]
    {Start Time} & {End Time} & {model} & {$\alpha$} & {$\beta$} & {$E_{\text{p}}$} & {Flux} & {cstat/dof}& {BIC} & {Best model} \\
    {(s)} & {(s)} & {} & {} & {} & {(keV)} & {$10^{-5} \text{erg} \cdot \text{cm}^{-2} \cdot \text{s}^{-1} $} & {} & {} & {}\\
\endhead
\midrule[0.8pt]
\endlastfoot
\multirow{2}{*}{500.0} & \multirow{2}{*}{500.5} & Band & -1.64$^{+0.16}_{-0.05}$ & -4.68$^{+2.62}_{-2.90}$ & 596.79$^{+325.95}_{-347.16}$ & 6.23$^{+0.82}_{-0.86}$ & 47.27/51.00 & 63.00 & \multirow{2}{*}{CPL} \\
  & & CPL & -1.67$^{+0.04}_{-0.05}$ & -- & 880.09$^{+573.65}_{-341.51}$ & 6.96$^{+0.94}_{-0.98}$ & 48.80/52.00 & 60.00 & \\
\hline
\multirow{2}{*}{500.5} & \multirow{2}{*}{501.0} & Band & -1.62$^{+0.04}_{-0.04}$ & -5.12$^{+2.36}_{-2.60}$ & 881.52$^{+334.49}_{-257.17}$ & 7.70$^{+0.81}_{-0.84}$ & 56.98/51.00 & 73.00 & \multirow{2}{*}{CPL} \\
  & & CPL & -1.64$^{+0.04}_{-0.04}$ & -- & 1252.17$^{+729.94}_{-469.83}$ & 8.37$^{+1.03}_{-1.10}$ & 57.44/52.00 & 69.00 & \\
\hline
\multirow{2}{*}{501.0} & \multirow{2}{*}{501.4} & Band & -1.62$^{+0.04}_{-0.04}$ & -5.23$^{+2.21}_{-2.62}$ & 603.83$^{+257.44}_{-237.66}$ & 8.03$^{+0.88}_{-1.10}$ & 55.14/51.00 & 71.00 & \multirow{2}{*}{Band} \\
  & & CPL & -1.64$^{+0.04}_{-0.05}$ & -- & 720.02$^{+349.90}_{-227.70}$ & 8.30$^{+1.05}_{-0.97}$ & 71.08/52.00 & 83.00 & \\
\hline
\multirow{2}{*}{501.4} & \multirow{2}{*}{502.8} & Band & -1.57$^{+0.04}_{-0.03}$ & -6.95$^{+2.82}_{-1.82}$ & 484.14$^{+119.34}_{-83.47}$ & 6.24$^{+0.50}_{-0.41}$ & 40.84/51.00 & 56.00 & \multirow{2}{*}{CPL} \\
  & & CPL & -1.58$^{+0.03}_{-0.03}$ & -- & 519.37$^{+129.85}_{-96.42}$ & 6.35$^{+0.53}_{-0.46}$ & 40.81/52.00 & 52.00 & \\
\hline
\multirow{2}{*}{502.8} & \multirow{2}{*}{503.2} & Band & -1.64$^{+0.06}_{-0.06}$ & -5.56$^{+2.34}_{-2.27}$ & 618.54$^{+308.60}_{-209.83}$ & 5.83$^{+0.96}_{-0.83}$ & 56.83/51.00 & 72.00 & \multirow{2}{*}{CPL} \\
  & & CPL & -1.68$^{+0.05}_{-0.06}$ & -- & 785.36$^{+629.25}_{-312.44}$ & 6.34$^{+1.01}_{-0.87}$ & 56.70/52.00 & 68.00 & \\
\hline
\multirow{2}{*}{503.2} & \multirow{2}{*}{503.3} & Band & -1.46$^{+0.19}_{-0.13}$ & -6.01$^{+2.82}_{-2.16}$ & 339.14$^{+419.82}_{-187.42}$ & 7.03$^{+2.12}_{-1.94}$ & 56.98/51.00 & 69.00 & \multirow{2}{*}{CPL} \\
  & & CPL & -1.57$^{+0.09}_{-0.16}$ & -- & 827.19$^{+557.77}_{-507.65}$ & 7.14$^{+2.08}_{-2.09}$ & 53.36/52.00 & 65.00 & \\
\hline
\multirow{2}{*}{503.3} & \multirow{2}{*}{503.5} & Band & -1.52$^{+0.07}_{-0.05}$ & -6.16$^{+2.57}_{-2.22}$ & 829.30$^{+353.06}_{-245.90}$ & 10.04$^{+1.51}_{-1.34}$ & 65.75/51.00 & 81.00 & \multirow{2}{*}{CPL} \\
  & & CPL & -1.53$^{+0.06}_{-0.06}$ & -- & 883.98$^{+456.56}_{-273.26}$ & 10.20$^{+1.65}_{-1.36}$ & 65.74/52.00 & 77.00 & \\
\hline
\multirow{2}{*}{503.5} & \multirow{2}{*}{503.8} & Band & -1.57$^{+0.19}_{-0.05}$ & -4.53$^{+2.40}_{-3.17}$ & 627.30$^{+360.59}_{-390.74}$ & 10.70$^{+1.39}_{-1.53}$ & 48.84/51.00 & 64.00 & \multirow{2}{*}{CPL} \\
  & & CPL & -1.61$^{+0.04}_{-0.05}$ & -- & 854.28$^{+520.92}_{-258.69}$ & 11.10$^{+1.73}_{-1.32}$ & 51.93/52.00 & 63.00 & \\
\hline
\multirow{2}{*}{503.8} & \multirow{2}{*}{504.0} & Band & -1.61$^{+0.08}_{-0.06}$ & -5.72$^{+2.29}_{-2.46}$ & 532.03$^{+357.74}_{-188.64}$ & 8.21$^{+1.92}_{-1.32}$ & 44.98/51.00 & 61.00 & \multirow{2}{*}{CPL} \\
  & & CPL & -1.63$^{+0.07}_{-0.09}$ & -- & 612.35$^{+577.26}_{-257.06}$ & 8.43$^{+1.79}_{-1.44}$ & 44.98/52.00 & 57.00 & \\
\hline
\multirow{2}{*}{504.0} & \multirow{2}{*}{505.0} & Band & -1.65$^{+0.03}_{-0.03}$ & -5.81$^{+2.59}_{-2.15}$ & 742.12$^{+247.46}_{-194.33}$ & 7.76$^{+0.68}_{-0.69}$ & 54.21/51.00 & 70.00 & \multirow{2}{*}{CPL} \\
  & & CPL & -1.65$^{+0.03}_{-0.03}$ & -- & 773.22$^{+326.61}_{-205.83}$ & 7.81$^{+0.82}_{-0.74}$ & 54.24/52.00 & 66.00 & \\
\hline
\multirow{2}{*}{505.0} & \multirow{2}{*}{505.9} & Band & -1.57$^{+0.04}_{-0.03}$ & -5.14$^{+2.45}_{-2.67}$ & 743.28$^{+254.43}_{-203.29}$ & 8.59$^{+0.85}_{-0.83}$ & 69.06/51.00 & 85.00 & \multirow{2}{*}{CPL} \\
  & & CPL & -1.58$^{+0.03}_{-0.03}$ & -- & 795.88$^{+240.05}_{-190.07}$ & 8.55$^{+0.79}_{-0.80}$ & 70.88/52.00 & 82.00 & \\
\hline
\multirow{2}{*}{505.9} & \multirow{2}{*}{506.1} & Band & -1.47$^{+0.06}_{-0.06}$ & -5.38$^{+2.17}_{-2.59}$ & 850.21$^{+416.16}_{-239.67}$ & 10.36$^{+1.75}_{-1.39}$ & 64.62/51.00 & 80.00 & \multirow{2}{*}{CPL} \\
  & & CPL & -1.51$^{+0.06}_{-0.07}$ & -- & 1023.24$^{+683.23}_{-353.11}$ & 10.85$^{+2.16}_{-1.66}$ & 64.58/52.00 & 76.00 & \\
\hline
\multirow{2}{*}{506.1} & \multirow{2}{*}{506.4} & Band & -1.57$^{+0.04}_{-0.04}$ & -6.12$^{+2.44}_{-2.43}$ & 941.96$^{+301.40}_{-227.96}$ & 11.83$^{+1.12}_{-1.15}$ & 39.65/51.00 & 45.00 & \multirow{2}{*}{CPL} \\
  & & CPL & -1.59$^{+0.04}_{-0.05}$ & -- & 1107.68$^{+595.86}_{-328.74}$ & 12.35$^{+1.68}_{-1.43}$ & 39.64/52.00 & 41.00 & \\
\hline
\multirow{2}{*}{506.4} & \multirow{2}{*}{506.7} & Band & -1.42$^{+0.19}_{-0.08}$ & -3.28$^{+1.11}_{-3.70}$ & 389.17$^{+255.69}_{-174.98}$ & 11.09$^{+1.65}_{-1.61}$ & 64.58/51.00 & 80.00 & \multirow{2}{*}{Band} \\
  & & CPL & -1.49$^{+0.05}_{-0.07}$ & -- & 598.35$^{+246.12}_{-191.91}$ & 11.39$^{+1.60}_{-1.66}$ & 69.77/52.00 & 81.00 & \\
\hline
\multirow{2}{*}{506.7} & \multirow{2}{*}{507.0} & Band & -1.54$^{+0.05}_{-0.05}$ & -5.94$^{+2.11}_{-2.13}$ & 534.96$^{+198.08}_{-115.91}$ & 9.99$^{+1.24}_{-0.96}$ & 61.89/51.00 & 77.00 & \multirow{2}{*}{CPL} \\
  & & CPL & -1.55$^{+0.05}_{-0.06}$ & -- & 561.75$^{+198.30}_{-145.66}$ & 10.10$^{+1.15}_{-1.10}$ & 61.89/52.00 & 73.00 & \\
\hline
\multirow{2}{*}{507.0} & \multirow{2}{*}{507.3} & Band & -1.51$^{+0.12}_{-0.05}$ & -5.03$^{+2.77}_{-2.59}$ & 660.15$^{+244.66}_{-319.70}$ & 14.13$^{+1.61}_{-1.45}$ & 48.66/51.00 & 64.00 & \multirow{2}{*}{CPL} \\
  & & CPL & -1.53$^{+0.04}_{-0.05}$ & -- & 734.70$^{+236.19}_{-168.83}$ & 14.02$^{+1.53}_{-1.34}$ & 50.48/52.00 & 62.00 & \\
\hline
\multirow{2}{*}{507.3} & \multirow{2}{*}{507.9} & Band & -1.55$^{+0.03}_{-0.03}$ & -6.28$^{+2.42}_{-1.91}$ & 602.50$^{+116.90}_{-84.09}$ & 12.22$^{+0.81}_{-0.72}$ & 67.08/51.00 & 83.00 & \multirow{2}{*}{CPL} \\
  & & CPL & -1.55$^{+0.03}_{-0.03}$ & -- & 609.55$^{+110.05}_{-87.98}$ & 12.18$^{+0.79}_{-0.71}$ & 67.00/52.00 & 79.00 & \\
\hline
\multirow{2}{*}{507.9} & \multirow{2}{*}{508.3} & Band & -1.56$^{+0.05}_{-0.04}$ & -5.63$^{+2.60}_{-2.31}$ & 771.47$^{+318.05}_{-240.32}$ & 10.76$^{+1.32}_{-1.27}$ & 48.24/51.00 & 64.00 & \multirow{2}{*}{CPL} \\
  & & CPL & -1.57$^{+0.04}_{-0.05}$ & -- & 810.61$^{+443.06}_{-234.76}$ & 10.77$^{+1.64}_{-1.29}$ & 48.54/52.00 & 60.00 & \\
\hline
\multirow{1}{*}{508.3} & \multirow{1}{*}{508.4} & Band & -1.47$^{+0.07}_{-0.06}$ & -6.14$^{+2.69}_{-2.04}$ & 1202.95$^{+454.60}_{-381.87}$ & 16.09$^{+2.51}_{-2.52}$ & 42.97/51.00 & 59.00 & \multirow{1}{*}{CPL} \\
\hline
 \multirow{1}{*}{508.3} & \multirow{1}{*}{508.4} & CPL & -1.51$^{+0.06}_{-0.06}$ & -- & 1705.49$^{+1008.13}_{-675.01}$ & 17.64$^{+2.96}_{-2.99}$ & 42.97/52.00 & 54.00 & \multirow{1}{*}{CPL}\\
\hline
\multirow{2}{*}{508.4} & \multirow{2}{*}{508.5} & Band & -1.27$^{+0.46}_{-0.22}$ & -2.77$^{+0.94}_{-3.91}$ & 311.11$^{+587.63}_{-207.99}$ & 15.29$^{+3.54}_{-4.04}$ & 48.96/51.00 & 64.00 & \multirow{2}{*}{Band} \\
 & & CPL & -1.51$^{+0.06}_{-0.08}$ & -- & 1061.17$^{+973.51}_{-479.23}$ & 16.41$^{+4.06}_{-3.45}$ & 60.15/52.00 & 72.00 & \\
\hline
\multirow{2}{*}{508.5} & \multirow{2}{*}{508.6} & Band & -1.50$^{+0.08}_{-0.06}$ & -6.31$^{+2.89}_{-2.34}$ & 801.04$^{+471.72}_{-269.36}$ & 17.36$^{+3.34}_{-2.59}$ & 40.49/51.00 & 56.00 & \multirow{2}{*}{CPL} \\
 &  & CPL & -1.54$^{+0.06}_{-0.06}$ & -- & 1144.90$^{+840.56}_{-452.51}$ & 19.39$^{+3.59}_{-3.32}$ & 40.48/52.00 & 52.00 & \\
\hline
\multirow{2}{*}{508.6} & \multirow{2}{*}{508.7} & Band & -1.17$^{+0.11}_{-0.07}$ & -5.27$^{+2.58}_{-2.47}$ & 487.66$^{+150.92}_{-132.29}$ & 19.23$^{+3.08}_{-2.41}$ & 63.31/51.00 & 79.00 & \multirow{2}{*}{CPL} \\
  & & CPL & -1.23$^{+0.08}_{-0.07}$ & -- & 570.12$^{+190.44}_{-117.97}$ & 19.82$^{+3.07}_{-2.42}$ & 65.43/52.00 & 77.00 & \\
\hline
\multirow{2}{*}{508.7} & \multirow{2}{*}{508.9} & Band & -1.44$^{+0.02}_{-0.03}$ & -3.42$^{+1.48}_{-3.56}$ & 1325.19$^{+532.44}_{-426.64}$ & 34.93$^{+3.03}_{-2.93}$ & 67.30/51.00 & 83.00 & \multirow{2}{*}{Band} \\
  & & CPL & -1.47$^{+0.03}_{-0.03}$ & -- & 1689.96$^{+569.54}_{-375.74}$ & 35.52$^{+3.71}_{-3.33}$ & 73.97/52.00 & 85.00 & \\
\hline
\multirow{2}{*}{508.9} & \multirow{2}{*}{509.3} & Band & -1.46$^{+0.03}_{-0.03}$ & -5.91$^{+2.36}_{-2.12}$ & 1082.67$^{+168.10}_{-154.93}$ & 27.02$^{+1.77}_{-1.63}$ & 61.99/51.00 & 78.00 & \multirow{2}{*}{CPL} \\
  & & CPL & -1.47$^{+0.02}_{-0.03}$ & -- & 1101.48$^{+173.62}_{-152.64}$ & 27.00$^{+1.68}_{-1.60}$ & 61.78/52.00 & 73.00 & \\
\hline
\multirow{2}{*}{509.3} & \multirow{2}{*}{509.5} & Band & -1.42$^{+0.10}_{-0.06}$ & -3.90$^{+1.75}_{-3.45}$ & 749.42$^{+301.90}_{-352.14}$ & 20.75$^{+2.39}_{-2.31}$ & 52.26/51.00 & 68.00 & \multirow{2}{*}{CPL} \\
  & & CPL & -1.45$^{+0.04}_{-0.05}$ & -- & 894.64$^{+307.49}_{-210.581}$ & 20.65$^{+2.69}_{-2.26}$ & 55.31/52.00 & 67.00 & \\
\hline
\multirow{2}{*}{509.5} & \multirow{2}{*}{510.0} & Band & -1.48$^{+0.03}_{-0.03}$ & -6.18$^{+2.37}_{-2.26}$ & 1014.16$^{+164.61}_{-140.34}$ & 24.58$^{+1.47}_{-1.38}$ & 64.14/51.00 & 80.00 & \multirow{2}{*}{CPL} \\
  & & CPL & -1.48$^{+0.02}_{-0.02}$ & -- & 1036.04$^{+169.96}_{-134.07}$ & 24.66$^{+1.52}_{-1.38}$ & 63.68/52.00 & 75.00 & \\
\hline
\multirow{2}{*}{510.0} & \multirow{2}{*}{510.1} & Band & -1.32$^{+0.05}_{-0.05}$ & -6.12$^{+2.17}_{-2.31}$ & 1152.41$^{+271.70}_{-189.82}$ & 38.94$^{+4.19}_{-3.69}$ & 65.38/51.00 & 81.00 & \multirow{2}{*}{CPL} \\
  & & CPL & -1.33$^{+0.05}_{-0.05}$ & -- & 1187.95$^{+306.44}_{-217.92}$ & 38.96$^{+4.29}_{-3.76}$ & 65.36/52.00 & 77.00 & \\
\hline
\multirow{2}{*}{510.1} & \multirow{2}{*}{510.2} & Band & -1.36$^{+0.28}_{-0.06}$ & -2.90$^{+1.16}_{-4.24}$ & 1194.53$^{+578.07}_{-977.05}$ & 55.57$^{+6.62}_{-6.37}$ & 62.62/51.00 & 78.00 & \multirow{2}{*}{CPL} \\
  & & CPL & -1.42$^{+0.04}_{-0.04}$ & -- & 1731.08$^{+688.84}_{-422.05}$ & 55.73$^{+6.85}_{-6.13}$ & 62.96/52.00 & 74.00 & \\
\hline
\multirow{2}{*}{510.2} & \multirow{2}{*}{510.3} & Band & -1.39$^{+0.04}_{-0.03}$ & -5.73$^{+2.43}_{-2.54}$ & 1466.18$^{+280.96}_{-268.53}$ & 63.14$^{+5.10}_{-5.30}$ & 50.78/51.00 & 66.00 & \multirow{2}{*}{Band} \\
  & & CPL & -1.40$^{+0.04}_{-0.04}$ & -- & 1529.98$^{+400.19}_{-290.44}$ & 63.31$^{+6.30}_{-5.86}$ & 76.63/52.00 & 88.00 & \\
\hline
\multirow{2}{*}{510.3} & \multirow{2}{*}{510.4} & Band & -1.25$^{+0.04}_{-0.03}$ & -6.25$^{+2.35}_{-1.84}$ & 1455.48$^{+225.52}_{-179.52}$ & 83.55$^{+6.78}_{-6.02}$ & 62.99/51.00 & 79.00 & \multirow{2}{*}{CPL} \\
  & & CPL & -1.26$^{+0.04}_{-0.04}$ & -- & 1486.68$^{+243.40}_{-184.76}$ & 83.46$^{+6.49}_{-6.03}$ & 62.94/52.00 & 74.00 & \\
\hline
\multirow{2}{*}{510.4} & \multirow{2}{*}{510.7} & Band & -1.31$^{+0.02}_{-0.02}$ & -3.25$^{+0.86}_{-3.72}$ & 1383.45$^{+157.11}_{-147.20}$ & 95.00$^{+3.96}_{-4.15}$ & 82.32/51.00 & 98.00 & \multirow{2}{*}{Band} \\
  & & CPL & -1.31$^{+0.02}_{-0.02}$ & -- & 1478.08$^{+130.34}_{-117.05}$ & 93.94$^{+3.91}_{-3.68}$ & 87.13/52.00 & 99.00 & \\
\hline
\multirow{2}{*}{510.7} & \multirow{2}{*}{510.8} & Band & -1.34$^{+0.05}_{-0.04}$ & -4.73$^{+2.38}_{-2.83}$ & 1437.28$^{+304.02}_{-317.67}$ & 80.41$^{+6.21}_{-6.13}$ & 65.88/51.00 & 81.00 & \multirow{2}{*}{Band} \\
  & & CPL & -1.29$^{+0.01}_{-0.01}$ & -- & 1629.87$^{+154.13}_{-125.43}$ & 87.18$^{+3.47}_{-2.58}$ & 114.14/52.00 & 126.00 & \\
\hline
\multirow{2}{*}{510.8} & \multirow{2}{*}{510.9} & Band & -1.29$^{+0.10}_{-0.05}$ & -3.66$^{+1.63}_{-3.55}$ & 1217.67$^{+305.48}_{-570.24}$ & 65.47$^{+5.74}_{-5.04}$ & 44.06/51.00 & 60.00 & \multirow{2}{*}{CPL} \\
  & & CPL & -1.32$^{+0.04}_{-0.03}$ & -- & 1388.21$^{+264.75}_{-218.93}$ & 64.77$^{+5.59}_{-5.25}$ & 47.35/52.00 & 59.00 & \\
\hline
\multirow{2}{*}{510.9} & \multirow{2}{*}{511.3} & Band & -1.33$^{+0.03}_{-0.02}$ & -2.52$^{+0.29}_{-2.55}$ & 1057.03$^{+179.10}_{-162.83}$ & 55.78$^{+2.85}_{-2.80}$ & 63.17/51.00 & 79.00 & \multirow{2}{*}{Band} \\
  & & CPL & -1.35$^{+0.02}_{-0.02}$ & -- & 1235.27$^{+125.31}_{-123.98}$ & 54.21$^{+2.59}_{-2.53}$ & 69.48/52.00 & 81.00 & \\
\hline
\multirow{2}{*}{511.3} & \multirow{2}{*}{511.5} & Band & -1.31$^{+0.03}_{-0.03}$ & -6.51$^{+2.43}_{-2.07}$ & 1255.58$^{+162.12}_{-156.93}$ & 50.96$^{+3.10}_{-3.37}$ & 59.14/51.00 & 75.00 & \multirow{2}{*}{CPL} \\
  & & CPL & -1.32$^{+0.03}_{-0.03}$ & -- & 1256.76$^{+172.32}_{-149.05}$ & 50.54$^{+3.25}_{-3.14}$ & 58.92/52.00 & 71.00 & \\
\hline
\multirow{2}{*}{511.5} & \multirow{2}{*}{511.9} & Band & \multicolumn{6}{c}{Unconstrained} & \multirow{2}{*}{CPL} \\
  & & CPL & -1.35$^{+0.01}_{-0.05}$ & -- & 1130.27$^{+196.92}_{-164.31}$ & 42.70$^{+1.74}_{-1.35}$ & 135.10/60.00 & 147.00 & \\
\hline
\multirow{2}{*}{511.9} & \multirow{2}{*}{512.1} & Band & -1.19$^{+0.14}_{-0.14}$ & -1.95$^{+0.10}_{-0.69}$ & 383.97$^{+554.40}_{-131.68}$ & 37.61$^{+3.27}_{-2.94}$ & 63.73/51.00 & 79.00 & \multirow{2}{*}{Band} \\
  & & CPL & -1.40$^{+0.01}_{-0.01}$ & -- & 1146.27$^{+106.35}_{-109.67}$ & 34.95$^{+1.01}_{-1.10}$ & 78.33/52.00 & 90.00 & \\
\hline
\multirow{2}{*}{512.1} & \multirow{2}{*}{512.5} & Band & -1.36$^{+0.03}_{-0.03}$ & -2.47$^{+0.27}_{-1.77}$ & 900.22$^{+198.17}_{-164.18}$ & 31.42$^{+2.06}_{-1.88}$ & 59.70/51.00 & 75.00 & \multirow{2}{*}{Band} \\
  & & CPL & -1.41$^{+0.01}_{-0.01}$ & -- & 1122.25$^{+103.47}_{-101.54}$ & 30.19$^{+0.67}_{-0.69}$ & 68.45/52.00 & 80.00 & \\
\hline
\multirow{2}{*}{512.5} & \multirow{2}{*}{512.8} & Band & -1.43$^{+0.04}_{-0.03}$ & -5.00$^{+2.48}_{-2.67}$ & 1263.36$^{+305.35}_{-271.37}$ & 27.49$^{+2.29}_{-2.10}$ & 45.27/51.00 & 61.00 & \multirow{2}{*}{CPL} \\
  & & CPL & -1.44$^{+0.03}_{-0.03}$ & -- & 1344.16$^{+300.21}_{-244.79}$ & 27.48$^{+2.29}_{-2.27}$ & 46.79/52.00 & 58.00 & \\
\hline
\multirow{2}{*}{512.8} & \multirow{2}{*}{513.2} & Band & -1.48$^{+0.03}_{-0.03}$ & -5.42$^{+2.26}_{-2.46}$ & 1365.78$^{+318.61}_{-259.07}$ & 23.45$^{+1.85}_{-1.89}$ & 52.20/51.00 & 68.00 & \multirow{2}{*}{CPL} \\
 &  & CPL & -1.49$^{+0.03}_{-0.03}$ & -- & 1405.33$^{+393.33}_{-258.43}$ & 23.31$^{+2.17}_{-1.74}$ & 52.28/52.00 & 64.00 & \\
\hline
\multirow{2}{*}{513.2} & \multirow{2}{*}{513.5} & Band & -1.43$^{+0.02}_{-0.03}$ & -5.98$^{+2.49}_{-2.09}$ & 824.53$^{+201.66}_{-148.04}$ & 17.20$^{+1.80}_{-1.49}$ & 58.10/51.00 & 74.00 & \multirow{2}{*}{CPL} \\
& & CPL & -1.41$^{+0.04}_{-0.05}$ & -- & 772.54$^{+196.15}_{-162.17}$ & 16.61$^{+1.72}_{-1.68}$ & 52.12/52.00 & 70.00 & \\
\hline
\multirow{2}{*}{513.5} & \multirow{2}{*}{514.0} & Band & -1.44$^{+0.05}_{-0.04}$ & -3.06$^{+0.95}_{-3.73}$ & 1044.47$^{+362.81}_{-318.34}$ & 18.70$^{+1.62}_{-1.52}$ & 77.29/51.00 & 94.00 & \multirow{2}{*}{CPL} \\
 &  & CPL & -1.46$^{+0.03}_{-0.03}$ & -- & 1278.66$^{+298.72}_{-253.89}$ & 18.65$^{+1.61}_{-1.64}$ & 81.33/52.00 & 93.00 & \\
\hline
\multirow{2}{*}{514.0} & \multirow{2}{*}{514.5} & Band & -1.44$^{+0.08}_{-0.05}$ & -2.22$^{+0.29}_{-3.87}$ & 769.91$^{+559.01}_{-363.84}$ & 16.15$^{+1.54}_{-1.49}$ & 48.77/51.00 & 64.00 & \multirow{2}{*}{Band} \\
  & & CPL & -1.48$^{+0.03}_{-0.04}$ & -- & 1201.05$^{+351.51}_{-276.88}$ & 15.71$^{+1.54}_{-1.56}$ & 56.68/52.00 & 68.00 & \\
\hline
\multirow{2}{*}{514.5} & \multirow{2}{*}{515.4} & Band & -1.52$^{+0.03}_{-0.02}$ & -4.64$^{+2.44}_{-2.99}$ & 1364.12$^{+256.26}_{-340.30}$ &14.87$^{+0.86}_{-0.93}$ & 40.52/51.00 & 56.00 & \multirow{2}{*}{CPL} \\
  & & CPL & -1.53$^{+0.03}_{-0.03}$ & -- & 1470.99$^{+354.74}_{-241.43}$ & 14.87$^{+1.07}_{-0.95}$ & 43.25/52.00 & 55.00 & \\
\hline
\multirow{2}{*}{515.4} & \multirow{2}{*}{516.3} & Band & -1.43$^{+0.21}_{-0.06}$ & -2.07$^{+0.25}_{-3.98}$ & 747.21$^{+468.94}_{-524.43}$ & 13.06$^{+1.07}_{-1.19}$ & 48.86/51.00 & 64.00 & \multirow{2}{*}{Band} \\
  & & CPL & -1.49$^{+0.03}_{-0.03}$ & -- & 1227.85$^{+309.55}_{-226.02}$ & 12.40$^{+1.08}_{-0.97}$ & 54.97/52.00 & 66.00 & \\
\hline
\multirow{2}{*}{516.3} & \multirow{2}{*}{517.9} & Band & -1.46$^{+0.05}_{-0.03}$ & -2.05$^{+0.16}_{-0.64}$ & 785.64$^{+338.81}_{-309.32}$ & 12.27$^{+0.85}_{-0.87}$ & 52.14/51.00 & 68.00 & \multirow{2}{*}{Band} \\
  & & CPL & -1.50$^{+0.02}_{-0.02}$ & -- & 1264.79$^{+249.13}_{-205.80}$ & 11.66$^{+0.80}_{-0.80}$ & 62.16/52.00 & 74.00 & \\
\hline
\multirow{2}{*}{517.9} & \multirow{2}{*}{518.9} & Band & -1.41$^{+0.10}_{-0.12}$ & -1.90$^{+0.06}_{-1.99}$ & 313.75$^{+923.28}_{-99.49}$ & 10.81$^{+0.79}_{-0.65}$ & 60.07/51.00 & 76.00 & \multirow{2}{*}{CPL} \\
  & & CPL & -1.55$^{+0.03}_{-0.02}$ & -- & 1381.84$^{+369.02}_{-261.40}$ & 10.74$^{+0.82}_{-0.77}$ & 61.76/52.00 & 73.00 & \\
\hline
\multirow{2}{*}{518.9} & \multirow{2}{*}{519.3} & Band & -1.55$^{+0.09}_{-0.07}$ & -5.37$^{+2.73}_{-2.53}$ & 536.67$^{+421.82}_{-221.68}$ & 6.64$^{+1.39}_{-1.19}$ & 55.45/51.00 & 71.00 & \multirow{2}{*}{CPL} \\
  & & CPL & -1.59$^{+0.06}_{-0.07}$ & -- & 747.90$^{+744.84}_{-311.68}$ & 7.13$^{+1.61}_{-1.27}$ & 58.03/52.00 & 70.00 & \\
\hline
\multirow{2}{*}{519.3} & \multirow{2}{*}{520.0} & Band & -1.50$^{+0.14}_{-0.04}$ & -3.62$^{+1.73}_{-3.67}$ & 1147.26$^{+426.32}_{-869.16}$ & 10.07$^{+0.88}_{-0.90}$ & 67.24/51.00 & 83.00 & \multirow{2}{*}{CPL} \\
  & & CPL & -1.54$^{+0.03}_{-0.03}$ & -- & 1607.51$^{+665.52}_{-446.24}$ & 10.46$^{+1.15}_{-1.14}$ & 70.76/52.00 & 82.00 & \\
\hline
%\end{deluxetable}
\end{longtable}

During the spectral fitting process, the CPL model was employed for the majority of time intervals, while the Band function was only applied to a few selected periods. All time intervals are relative to the trigger time $T_0$.

%\section{Spectra fitting results}

%\section{Appendix information}

%\section{Author publication charges} \label{sec:pubcharge}



\end{document}